\documentclass[aps,prl,twocolumn,superscriptaddress,groupedaddress,preprintnumbers]{revtex4-1} 
\usepackage[totalheight = 23cm, totalwidth = 17cm]{geometry}
\usepackage{amssymb,amsmath,amsfonts,amsbsy}
\usepackage{bm}
\usepackage{graphicx}  
\usepackage{dcolumn}   
\usepackage{bm}        
\usepackage{amssymb}   
\usepackage{amsmath}
\usepackage{mathtools}
\usepackage{cases}
\usepackage{mathrsfs}
\usepackage{color}
\usepackage{enumitem}
\makeatletter
\newcommand*\bigcdot{\mathpalette\bigcdot@{.5}}
\newcommand*\bigcdot@[2]{\mathbin{\vcenter{\hbox{\scalebox{#2}{$\m@th#1\bullet$}}}}}
\makeatother

\newcommand{\calR}{{\cal R}}
\newcommand{\be}{\begin{equation}}
\newcommand{\ee}{\end{equation}}
\newcommand{\bea}{\begin{eqnarray}}
\newcommand{\eea}{\end{eqnarray}}
\newcommand{\nn}{\nonumber}

\newcommand{\vf}{\varphi}

\newcommand*{\dif}{\mathop{}\!\mathrm{d}}

\begin{document}
	\preprint{IPMU22-0060}
	\preprint{YITP-22-144}
	
		\title{Logarithmic Duality of the Curvature Perturbation}
	\author{Shi Pi${}^{a,b,c}$} \email{shi.pi@itp.ac.cn} 
	\author{Misao Sasaki${}^{c,d,e}$}
		\affiliation{
		$^{a}$ CAS Key Laboratory of Theoretical Physics, Institute of Theoretical Physics, Chinese Academy of Sciences, Beijing 100190, China \\
		$^{b}$ Center for High Energy Physics, Peking University, Beijing 100871, China\\
		$^{c}$ Kavli Institute for the Physics and Mathematics of the Universe (WPI), The University of Tokyo, Kashiwa, Chiba 277-8583, Japan\\
		$^{d}$ Center for Gravitational Physics and Quantum Information,
  Yukawa Institute for Theoretical Physics, Kyoto University, Kyoto 606-8502, Japan\\
		$^{e}$ Leung Center for Cosmology and Particle Astrophysics,\\
  National Taiwan University, Taipei 10617}
	\date{\today}
	\begin{abstract}
We study the comoving curvature perturbation $\mathcal{R}$ in the single-field inflation models whose potential can be approximated by a piecewise quadratic potential $V(\varphi)$ by using the $\delta N$ formalism. We find a general formula for $\mathcal{R}(\delta\varphi, \delta\pi)$, consisting of a sum of logarithmic functions of the field perturbation $\delta\varphi$ and the velocity perturbation $\delta\pi$ at the point of interest, as well as of $\delta\pi_*$ at the boundaries of each quadratic piece, which are functions of ($\delta\varphi, \delta\pi$) through the equation of motion. Each logarithmic expression has an equivalent dual expression, due to the second-order nature of the equation of motion for $\varphi$. We also clarify the condition under which $\mathcal{R}(\delta\varphi, \delta\pi)$ reduces to a single logarithm, which yields either the renowned ``exponential tail'' of the probability distribution function of $\mathcal{R}$ or a Gumbel-distribution-like tail.
\end{abstract}
\maketitle

\textit{Introduction.}---The primordial curvature perturbation on comoving slices $\mathcal{R}$ originates from the quantum fluctuations of the inflaton $\vf$ during inflation \cite{Brout:1977ix,Guth:1980zm,Starobinsky:1980te,Mukhanov:1981xt,Linde:1981mu,Albrecht:1982wi}. In linear perturbation theory,
$\mathcal{R}=-(H/\dot\vf)\delta\vf$ \cite{Mukhanov:1985rz,Sasaki:1986hm}, where $\delta\vf$ is the field perturbation on spatially-flat slices.
The observed curvature perturbation is Gaussian, and has a nearly scale-variant power spectrum $\mathcal{P_R}$ of order $10^{-9}$ on scales $\gtrsim10$ Mpc~\cite{Planck:2018jri,DES:2021wwk}. 
However, on small scales, $\mathcal{P_R}$ is not well constrained due to nonlinear astrophysical processes. Thus $\mathcal{P_R}$ might be much enhanced on small scales, which could lead to interesting phenomena, for instance,  the formation of primordial black holes (PBHs) \cite{Zeldovich:1967lct,Hawking:1971ei,Carr:1974nx,Meszaros:1974tb,Carr:1975qj,Khlopov:1985jw} and induced gravitational waves (GWs) \cite{Matarrese:1992rp,Matarrese:1993zf,Matarrese:1997ay,Noh:2004bc,Carbone:2004iv,Nakamura:2004rm,Ananda:2006af,Baumann:2007zm}. In such models, the enhanced power spectrum are often accompanied by $\mathcal{R}$ in the form of nonlinear functions of $\delta\vf$, which can be calculated by the $\delta N$ formalism.

The $\delta N$ formalism \cite{Sasaki:1995aw,Wands:2000dp,Lyth:2004gb} connects the comoving curvature perturbation $\calR$ to the field perturbation $\delta\vf$ and the velocity perturbation $\delta\pi$, which are quantum fluctuations evaluated on spatially-flat slices on superhorizon scales. As the Hubble patches separated by superhorizon scales can be treated as casually disconnected ``separate universes'', the local expansion rate in such a patch is randomly distributed according to its probability distribution function (PDF). Along a trajectory starting from an initial spatially flat slice to a final comoving slice, the difference between its total expansion, or the $e$-folding number, and the fiducial total expansion equals to the curvature perturbation on the final comoving slice in this patch, i.e. $\calR=\delta N(\delta\vf, \delta\pi)$.
For slow-roll inflation, as the non-Gaussianity is small, and $\delta\pi$ is negligible, we have the perturbative series $\calR=(\partial N/\partial\vf)\delta\vf+(1/2)(\partial^2N/\partial\vf^2)\delta\vf^2+\cdots$ \cite{Lyth:2005fi}.

However, even a small non-Gaussianity can significantly change the tail of the PDF of $\calR$, thus alters, for instance, the PBH mass function greatly, as the formation of compact objects like PBHs depends sensitively on the tail of the PDF of $\calR$ \cite{Young:2013oia,Young:2015cyn,Atal:2018neu,Passaglia:2018ixg,Yoo:2019pma,Kehagias:2019eil,Mahbub:2020row,Riccardi:2021rlf,Davies:2021loj,Young:2022phe,Escriva:2022pnz,Matsubara:2022nbr}. 
Recently, it was discovered that a fully nonlinear logarithmic relation $\calR(\delta\vf)=-(1/\lambda)\ln\big[1+\mathcal{O}(\delta\vf)\big]$ can give a non-Gaussianity of $\mathcal{O}(1)$ in, e.g., the ultra-slow-roll (USR) inflation \cite{Cai:2018dkf,Biagetti:2018pjj}, inflation near a bump \cite{Atal:2019erb,Atal:2019cdz}, the curvaton scenario \cite{Pi:2021dft}, inflation with a step-up potential \cite{Cai:2021zsp,Cai:2022erk}, etc. 
This logarithmic relation generates an ``exponential tail'' of the PDF $P(\calR)\sim\exp(-\lambda\calR)$, similar to what is found in the stochastic approach \cite{Vennin:2015hra,Pattison:2017mbe,Ezquiaga:2019ftu,Vennin:2020kng,Figueroa:2020jkf,Pattison:2021oen,Figueroa:2021zah,Animali:2022otk}. 
It seems the logarithmic relation is quite common among many inflationary models, but its origin has not been clarified. Besides, the coefficients as well as the arguments of the logarithms are different for different models. Therefore it is worth investigating the mechanism of generating such logarithmic relations or exponential tails, and how their detailed forms depend on models.

\begin{figure}[htbp]
\begin{center}
\includegraphics[width=0.45\textwidth]{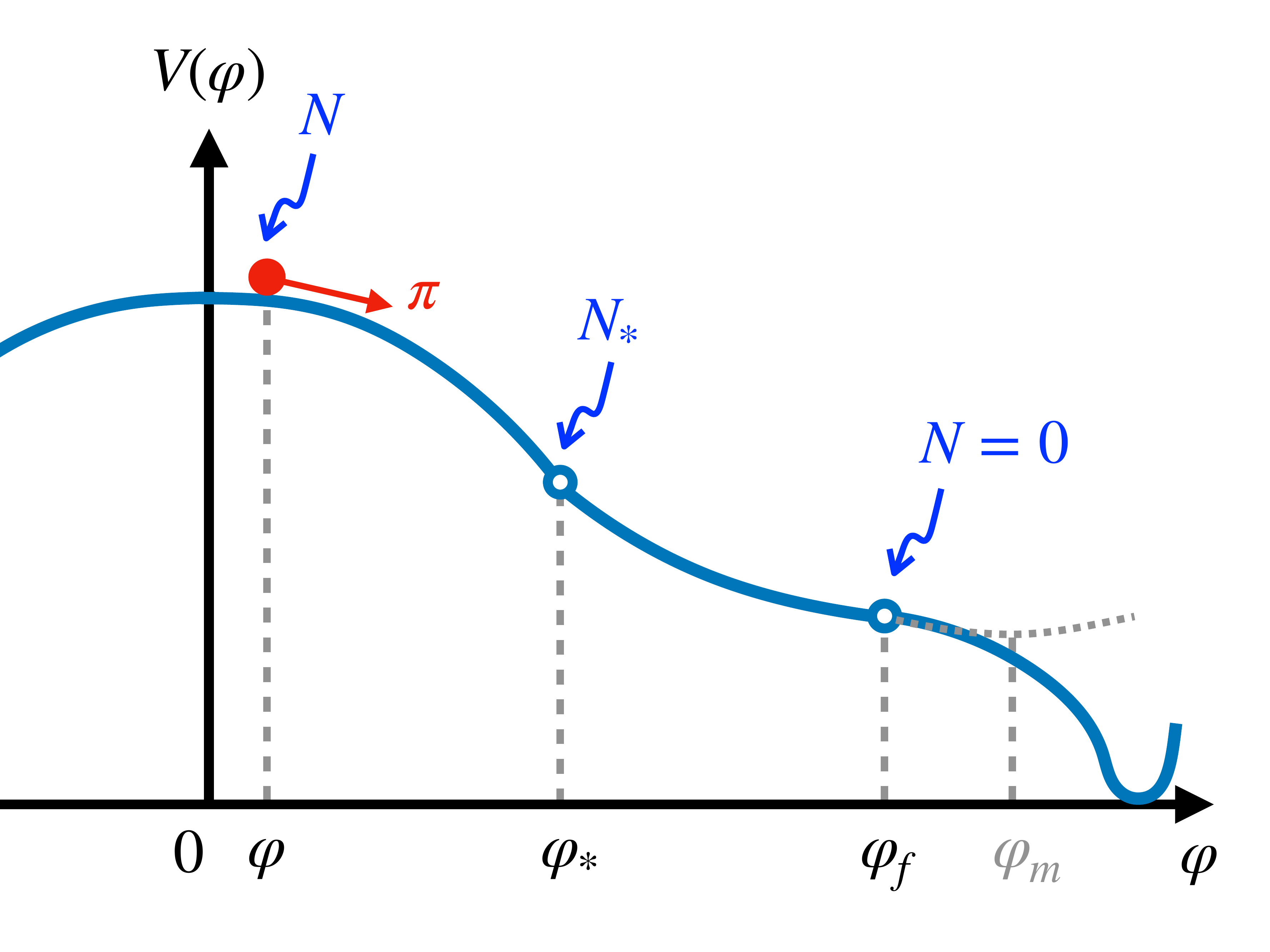}
\includegraphics[width=0.45\textwidth]{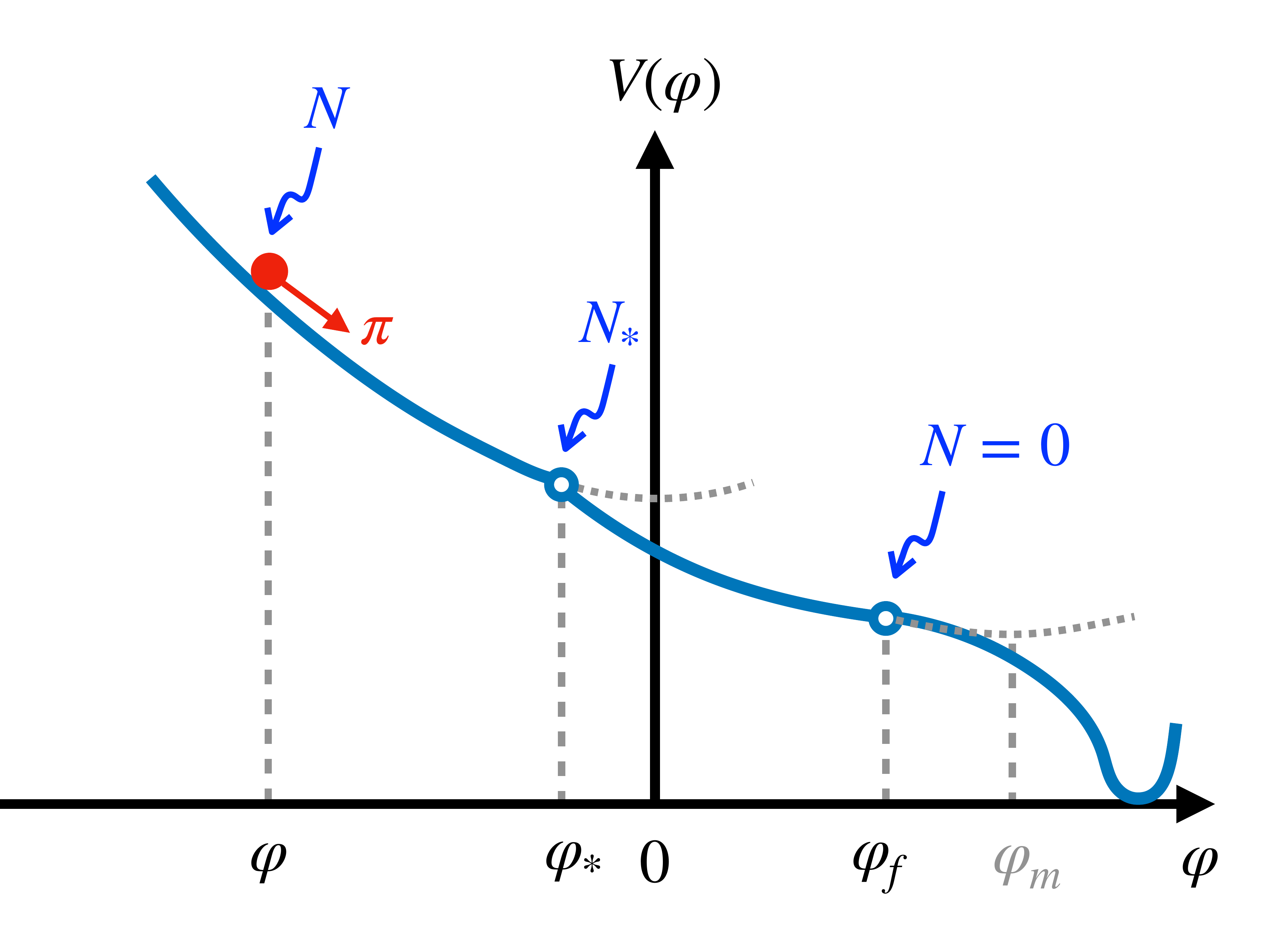}
\caption{Schematic pictures of $V(\vf)$, which glues two parabolas together at the junction point $\vf_*$, given by \eqref{V2} and \eqref{V1}. The origin is chosen at the local maximum (upper panel) or minimum (lower panel) of $V_1(\vf)$, for $m_1^2<0$ or $m_1^2>0$, respectively. At $\vf_*$, the potential is continuous, while its slope may not be. The $e$-folding number defined in \eqref{def:efold} is also labeled at $\vf$, $\vf_*$, and $\vf_f$.}
\label{f:potential}
\end{center}
\end{figure}

\textit{Logarithmic Duality.}---We consider a piecewise potential consisting of two parabolas:
\begin{align}\label{V2}
V_1(\vf<\vf_*)&=V_0+\frac{m_1^2}{2}\vf^2,\\\nn
V_2(\vf>\vf_*)&=V_0+\frac{m_1^2}{2}\vf_*^2
-\frac{m_2^2}{2}(\vf_*-\vf_m)^2
\\\label{V1}
&\quad+\frac{m_2^2}{2}\left(\vf-\vf_m\right)^2.
\end{align}
where $\vf_*$ is the junction point of the two potentials, and $\vf_m$ is the minimum of $V_2(\vf)$. For simplicity, we assume the origin of $\vf$ is at the maximum or minimum of $V_1(\vf)$, with $\vf_*>0$ or $\vf_*<0$, respectively. $V(\vf)$ is then a monotonic function around $\vf_*$. Inflation ends at $\vf_f$ in the second stage, i.e., $\vf_f>\vf_*$. A schematic figure of this piecewise potential is shown in Fig.~\ref{f:potential}. We consider a continuous $V(\vf)$, but there may be discontinuity in $V'(\vf)$ at $\vf_*$, unless
\be\label{def:vfm}
\vf_m=\left(1-\frac{m_1^2}{m_2^2}\right)\vf_*.
\ee
Although we only consider two segments here, an extension to more segments is straightforward, similar to what is done in Ref. \cite{Karam:2022nym}. 

The equation of motion of the inflaton field is $\ddot\vf+3H\dot\vf+\partial V/\partial\vf=0$.  It is convenient to define the $e$-folding number counted backward in time from the end of inflation,
\be\label{def:efold}
N=\int^{t_e}_t H\dif t\,,
\ee
and use it as the time variable. We assume that in the range of our interest, the first slow-roll parameter $\epsilon\equiv-\dot H/H^2$ is negligible, so that the Hubble parameter may be approximated by a constant value $3H^2\approx 8\pi G V_0$. The second slow-roll parameters $\eta\equiv m_1^2/(3H^2)$ and $\tilde\eta\equiv m_2^2/(3H^2)$ are constants, but we do not assume them to be small.
Then the equations of motion for $\vf$ are constant-coefficient second-order differential equations:
\begin{align}
\label{eom1}
&\frac{\dif^2\vf}{\dif N^2}-3\frac{\dif\vf}{\dif N}+3\eta\vf=0&\;(\vf\leq\vf_*),\\
\label{eom2}
&\frac{\dif^2\vf}{\dif N^2}-3\frac{\dif\vf}{\dif N}+3\tilde\eta(\vf-\vf_m)=0&\;(\vf>\vf_*).
\end{align}
Note that $N=0$ is at the end of inflation, while we assign $N=N_*$ at $\vf=\vf_*$. See Fig.~\ref{f:potential}.

Setting $\vf\propto e^{\lambda N}$, the characteristic root $\lambda$ of \eqref{eom1} is found as
\be\label{def:lambda}
\lambda^2-3\lambda+3\eta=0,\quad\Longrightarrow\quad \lambda_\pm=\frac{3\pm\sqrt{9-12\eta}}{2}.
\ee
For $\eta<3/4$, which we assume in this paper, we have $\lambda_-<\lambda_+$. 
The general solution of $\vf$ is
\be\label{vf1}
\vf(N)=c_+e^{\lambda_+(N-N_*)}+c_-e^{\lambda_-(N-N_*)},
\ee
where $c_\pm$ are constants. We define the field velocity as $\pi\equiv-\dif\vf/\dif N$, so that its sign is the same as $d\vf/dt$. 
Then
\be\label{pi1}
-\pi(N)=\lambda_+c_+e^{\lambda_+(N-N_*)}+\lambda_-c_-e^{\lambda_-(N-N_*)}.
\ee

The solution \eqref{vf1} and \eqref{pi1} are valid for $\vf\leq\vf_*$. Then the coefficients $c_\pm$ are
determined as
\be\label{cpm}
c_\pm=\mp\frac{\pi_*+\lambda_{\mp}\vf_*}{\lambda_+-\lambda_-},
\ee
where $\pi_*$ is the field velocity at $\vf_*$. Combining \eqref{vf1} and \eqref{pi1}, and using \eqref{cpm}, we have
\begin{align}
\label{com5}
\frac{\pi+\lambda_+\vf}{\pi_*+\lambda_+\vf_*}&=e^{\lambda_-(N-N_*)},\\
\label{com6}
\frac{\pi+\lambda_-\vf}{\pi_*+\lambda_-\vf_*}&=e^{\lambda_+(N-N_*)}.
\end{align}
These equations can be used to express the $e$-folding number $N-N_*$ in terms of $(\vf,\pi)$ and $\pi_*$, 
\begin{align}\label{N(phi)}
N-N_*=\frac{1}{\lambda_\pm}\ln\frac{\pi+\lambda_\mp\vf}{\pi_*+\lambda_\mp\vf_*}\,,
\end{align}
where $\pi_*$ is a function of $(\vf,\pi)$, determined by the equation combining \eqref{com5} and \eqref{com6},
\be\label{law}
\left(\frac{\pi+\lambda_+\vf}{\pi_*+\lambda_+\vf_*}\right)^{\lambda_+}
=\left(\frac{\pi+\lambda_-\vf}{\pi_*+\lambda_-\vf_*}\right)^{\lambda_-}.
\ee
We have two seemingly very different expressions for $N-N_*$ in \eqref{N(phi)}. But their equivalence can be easily shown by \eqref{law}. This is the origin of the logarithmic duality of the curvature perturbation.

The $\delta N$ formula can be obtained by subtracting the fiducial $e$-folding number \eqref{N(phi)} from a perturbed version with $N\to N+\delta N$, $N_*\to N_*+\delta N_*$, $\vf\to\vf+\delta\vf$, $\pi_*\to\pi_*+\delta\pi_*$, $\pi\to\pi+\delta\pi$,
\begin{align}\nn
&\delta(N-N_*)=\\\nn
&~\frac1{\lambda_\pm}\ln\left[1+\frac{\delta\pi+\lambda_\mp\delta\vf}{\pi+\lambda_\mp\vf}\right]-\frac1{\lambda_\pm}\ln\left[1+\frac{\delta\pi_*}{\pi_*+\lambda_\mp\vf_*}\right].\\\label{deltaN1}
\end{align}
The equivalence of the upper- and lower-sign formulas of \eqref{deltaN1} is guaranteed by taking the perturbation of \eqref{law},
\begin{align}\nn
&\quad\left(1+\frac{\delta\pi_*}{\pi_*+\lambda_+\vf_*}\right)^{-\lambda_+}\left(1+\frac{\delta\pi_*}{\pi_*+\lambda_-\vf_*}\right)^{\lambda_-}\\\label{law3}
&=\left(1+\frac{\delta\pi+\lambda_-\delta\vf}{\pi+\lambda_-\vf}\right)^{\lambda_-}\left(1+\frac{\delta\pi+\lambda_+\delta\vf}{\pi+\lambda_+\vf}\right)^{-\lambda_+}\,,
\end{align}
which also determines $\delta\pi_*$ as a function of ($\delta\vf, \delta\pi$) at an earlier stage. 


For  $\vf>\vf_*$, the equation of motion is given by \eqref{eom2}. 
Introducing $\tilde\vf\equiv\vf-\vf_m$, it becomes exactly in the same form as \eqref{eom1} 
with tilded $\vf$ and $\eta$.
With the junction point $\tilde\vf_*=\vf_*-\vf_m$ where $N=N_*$, and the endpoint $\tilde\vf_f=\vf_f-\vf_m$ where $N=0$,
in parallel with the previous discussion, we can calculate $\delta N_*$ of the second stage.
The resulting expression for the total $\delta N$ is
\begin{align}
\mathcal{R}\equiv\delta N=\delta(N-N_*)+\delta N_*\,,\label{main} 
\end{align}
where $\delta(N-N_*)$ is given by \eqref{deltaN1}, and $\delta N_*$ by
\begin{align}
&\delta N_*=
\nn\\
&~\frac1{\tilde\lambda_\pm}\ln\left[1+\frac{\delta\pi_*}{\pi_*+\tilde\lambda_\mp\tilde\vf_*}\right]
-\frac1{\tilde\lambda_{\pm}}\ln\left[1+\frac{\delta\pi_f}{\pi_f+\tilde\lambda_\mp\tilde\vf_f}\right]\,.
\label{deltaN2}
\end{align}
with $\tilde\lambda_\pm$ being the characteristic roots given by \eqref{def:lambda} with $\eta\to\tilde\eta$. 
\eqref{main} tells us that the curvature perturbation is the sum of logarithms of ($\delta\vf, \delta\pi$), as well as $\delta\pi$ at the junction ($\delta\pi_*$) and at the endpoint ($\delta\pi_f$), where $\delta\pi_f$ is a function of $\delta\pi_*$, via
 \begin{align}\nn
&\quad\left(1+\frac{\delta\pi_f}{\pi_f+\tilde\lambda_+\tilde\vf_f}\right)^{-\tilde\lambda_+}\left(1+\frac{\delta\pi_f}{\pi_f+\tilde\lambda_-\tilde\vf_f}\right)^{\tilde\lambda_-}\\\label{law4}
&=\left(1+\frac{\delta\pi_*}{\pi_*+\tilde\lambda_-\tilde\vf_*}\right)^{\tilde\lambda_-}\left(1+\frac{\delta\pi_*}{\pi_*+\tilde\lambda_+\tilde\vf_*}\right)^{-\tilde\lambda_+}\,.
\end{align}
Note that $\delta\pi_*$ is a function of ($\delta\vf, \delta\pi$) via \eqref{law3}. When evaluating \eqref{main}, the upper or lower signs in \eqref{deltaN1} and \eqref{deltaN2} can be chosen independently, of which the equivalence is guaranteed by \eqref{law3} and \eqref{law4}, respectively. We call this equivalence the \textit{logarithmic duality}, and this is the main result of our paper.

The main formula \eqref{main} together with \eqref{deltaN1} and \eqref{deltaN2} contains functions $\delta\pi_*(\delta\vf,\delta\pi)$ and $\delta\pi_f(\delta\vf,\delta\pi)$, which are determined by \eqref{law3} and \eqref{law4}. In general they can only be solved numerically.
However, \eqref{main} can be simplified greatly if the inflaton is already in the attractor regime at the boundaries.
Except for the degenerate limit $\lambda_+=\lambda_-=3/2$, we have $\lambda_+>\lambda_-$, hence the attractor solution is $\pi=-\lambda_-\vf$. 
Depending on the initial condition, the inflaton may already be in the attractor regime at $\vf_*$. 
If so, the second factor on the left hand side of \eqref{law3} is much larger than the first one. We can approximately solve for $\delta\pi_*$ to obtain
\be\label{law5}
1+\frac{\delta\pi_*}{\pi_*+\lambda_-\vf_*}
\approx
\left[1+\frac{\delta\pi+\lambda_-\delta\vf}{\pi+\lambda_-\vf}\right]\left[1+\frac{\delta\pi+\lambda_+\delta\vf}{\pi+\lambda_+\vf}\right]^{-\frac{\lambda_+}{\lambda_-}}.
\ee
Similarly, if the inflaton is in the attractor regime at the end of inflation, \eqref{law4} becomes
\be\label{law6}
1+\frac{\delta\pi_f}{\pi_f+\tilde\lambda_-\tilde\vf_f}
\approx\left[1+\frac{\delta\pi_*}{\pi_*+\tilde\lambda_-\tilde\vf_*}\right]\left[1+\frac{\delta\pi_*}{\pi_*+\tilde\lambda_+\tilde\vf_*}\right]^{-\frac{\tilde\lambda_+}{\tilde\lambda_-}}.
\ee
Substituting \eqref{law5} and \eqref{law6} into the upper-sign formulas of \eqref{deltaN1} and \eqref{deltaN2}, respectively, we find that the first terms in both expressions for $\delta(N-N_*)$ and $\delta N_*$ are canceled, leaving the lower-sign formulas without the contributions at the junction and the endpoint.
Summing up the resulting expressions, we obtain
\be\label{R1}
\mathcal{R}\approx\frac1{\lambda_-}\ln\left(1+\frac{\delta\pi+\lambda_+\delta\vf}{\pi+\lambda_+\vf}\right)+\frac1{\tilde\lambda_-}\ln\left(1+\frac{\delta\pi_*}{\pi_*+\tilde\lambda_+\tilde\vf_*}\right).
\ee

Apparently \eqref{R1} cannot be used when either $\lambda_-$ or $\tilde\lambda_-$ is zero, i.e. the USR case.  
During the USR stage the inflaton cannot be in the attractor regime and
 we have to use the upper-sign formula of \eqref{deltaN1} or \eqref{deltaN2}. 
For example, assuming the first stage is USR and the second stage ends in the attractor regime, we obtain
\be\label{R2}
\mathcal{R}\approx-\frac13\ln\left(1+\frac{\delta\pi_*}{\pi_*}\right)+\frac1{\tilde\lambda_-}\ln\left(1+\frac{\delta\pi_*}{\pi_*+\tilde\lambda_+\tilde\vf_*}\right),
\ee
where $\delta\pi_*$ is expressed in terms of $\delta\vf$ via \eqref{law3}, 
which now takes the form of a simple conservation law \cite{Namjoo:2012aa},
\be\label{law2}
\pi+3\vf=\pi_*+3\vf_*, \quad\Longrightarrow\quad \delta\pi_*=3\delta\vf\,.
\ee

Similarly, if the the inflaton is in the attractor regime when it reaches $\vf_*$, and the second stage is USR, we have
\be\label{R3}
\mathcal{R}\approx\frac1{\lambda_-}\ln\left(1+\frac{\delta\pi+\lambda_+\delta\vf}{\pi+\lambda_+\vf}\right)-\frac13\ln\left(1+\frac{\delta\pi_f}{\pi_f}\right).
\ee
As \eqref{law4} gives $\delta\pi_f=\delta\pi_*\approx-\lambda_-\delta\vf_*=0$, the second term is always negligible. 
We emphasize that in general once the inflaton is in the attractor regime, the trajectory in the later stages is unique. Therefore whatever feature the potential has in the following stage, it does not contribute to $\delta N$. 

\begin{figure}[htbp]
\begin{center}
\includegraphics[width=0.45\textwidth]{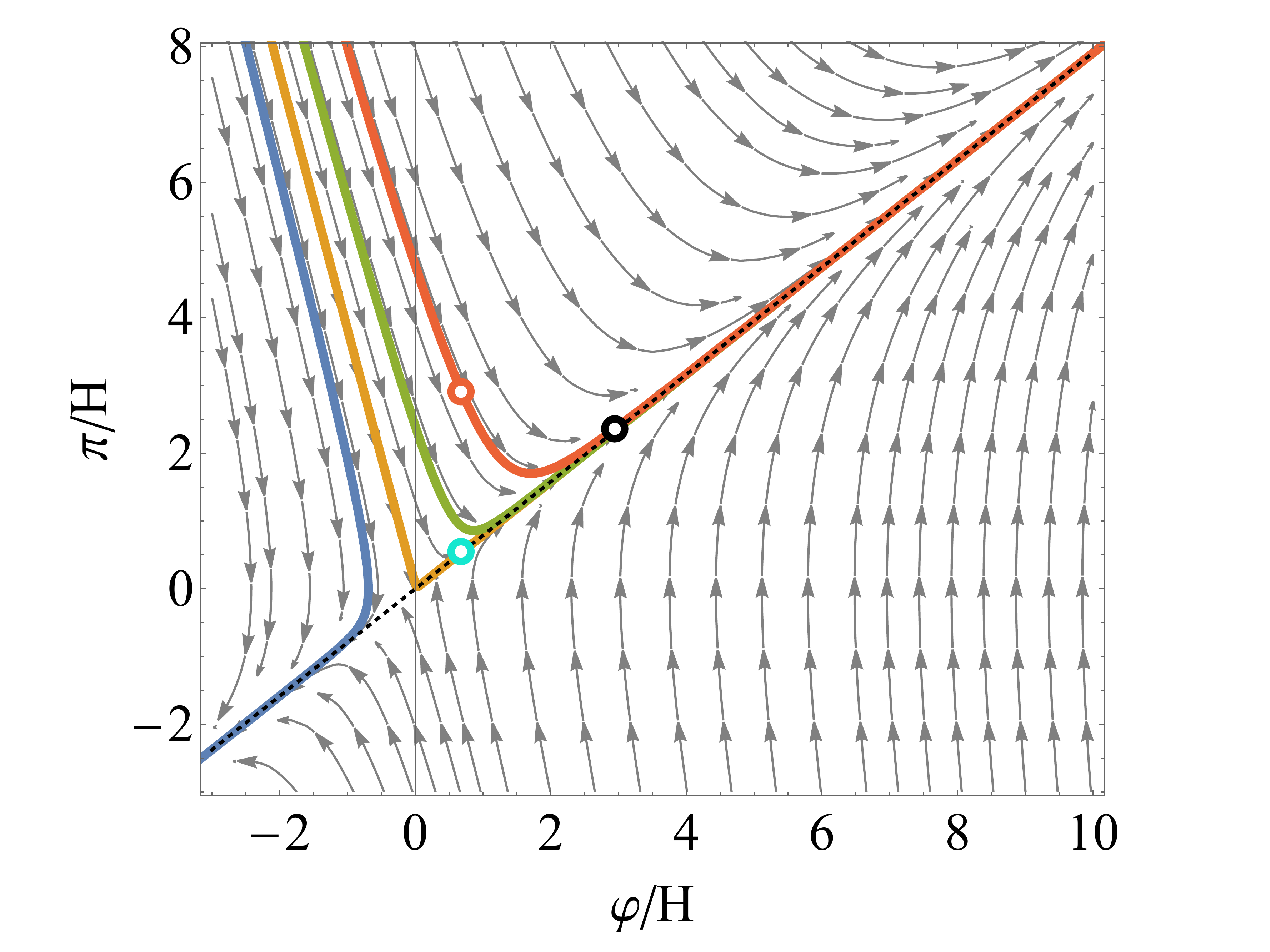}
\caption{Phase portrait of the equation of motion \eqref{eom1} near a bump with $\eta=-1$. The initial conditions for the blue, orange, green, and red curves are $\pi_i=10,\, 11.4,\, 13,\, 15$ at $\vf_i=-3$ (in the unit of $H$), respectively. The diagonal dotted line is the attractor solution $\pi+\lambda_-\vf=0$. According to the conservation of $(\delta\pi+\lambda_+\delta\vf)/(\pi+\lambda_+\vf)$, the superhorizon curvature perturbation \eqref{R4} at the red circle can be evaluated at a later time when $\vf$ is already in the attractor regime (the black circle), which can be approximately evaluated at an earlier moment in the attractor trajectory (the green circle).}
\label{f:phase}
\end{center}
\end{figure}

Actually, by checking \eqref{R1} and \eqref{R3}, we see that except for an extremely fine-tuned case of $\pi_*+\tilde\lambda_+\tilde\vf_*\approx0$, the contribution from the second stage is always negligible, leaving 
\be\label{R4}
\mathcal{R}\approx\frac{1}{\lambda_{-}}\ln\left(1+\frac{\delta\pi+\lambda_+\delta\vf}{\pi+\lambda_+\vf}\right),
\ee
provided the inflaton is already in the attractor regime at $\vf_*$. We note that under the assumption of a quadratic potential, perturbations $(\delta\vf,~\delta\pi)$ follow exactly the same equations for $(\vf,~\pi)$, i.e. \eqref{com5} and \eqref{com6}. Therefore $(\delta\pi+\lambda_+\delta\vf)/(\pi+\lambda_+\vf)$ is time independent, which can be calculated at any moment even in the attractor regime \footnote{We thank Jaume Garriga for pointing this out.}. 
This implies that we can calculate this quantity as if it were in the attractor solution at $\varphi(N)$ for any $N$ as long as the scale is outside the horizon, as was shown in \cite{Leach:2001zf} at the linear level. Thanks to our new logarithmic formula, we now have a fully nonlinear version of it. Namely, we can replace $\pi$ and $\delta\pi$ in \eqref{R4} with $-\lambda_-\varphi$ and $-\lambda_-\delta\varphi$, respectively. As an explicit example, see Fig.\ref{f:phase} \footnote{In this figure, we only consider $\vf>0$. We will leave $\vf\leq0$ case for future work.}.



\textit{Application to special cases.}---Our formula leads to a complicated form of $\calR(\delta\vf, \delta\pi)$ in general, which can only be calculated numerically. However, in some interesting special cases, it is possible to obtain approximated results. Besides, we can seek for analytically solvable cases which have not been studied before by our new formula. 


The first example is the slow-roll inflation, $\lambda_-\approx\eta$ and $\lambda_+\approx3-\eta$ with $|\eta|\ll1$. In this case, the inflaton is deep in the attractor regime at $\vf_*$. This means all the boundary terms are negligible if we use \eqref{R4},
\begin{align}\label{Rslowroll}
\mathcal{R}
&\approx\frac1\eta\ln\left[1+\frac{3\delta\vf}{\pi+3\vf}\right]\approx\frac1\eta\ln\left(1-\eta\frac{\delta\vf}{\pi}\right).
\end{align} 
In the second step we use the slow-roll equation of motion $\pi\approx-\eta\vf$. As $\delta\vf/\vf\ll 1$, it can be expanded as perturbation series
\be\label{Rseries}
\mathcal{R}\approx\calR_g+\frac35f_\text{NL}\calR_g^2+\cdots,
\ee
which yields the standard slow-roll result $\calR_g=-\delta\vf/\pi$ with 
$f_\mathrm{NL}=-5\eta/6$ \cite{Lyth:2005fi}.

The second example is USR inflation, where $\lambda_-=0$ and $\lambda_+=3$, while inflation ends at $\vf_*$. Then only the first term in \eqref{R2} remains to 
give \cite{Cai:2018dkf,Biagetti:2021eep}
\be
\mathcal{R}=-\frac1{3}\ln\left(1+\frac{\delta\pi_*}{\pi_*}\right)
=-\frac13\ln\left[1+\frac{3\delta\vf}{\pi+3(\vf-\vf_*)}\right],
\label{USR}
\ee
where in the second step \eqref{law2} is used.

If the USR stage is followed by a slow-roll stage, we have $\lambda_-=0$, $\lambda_+=3$, 
$\tilde\lambda_-=\tilde\eta$, and $\tilde\lambda_+=3-\tilde\eta$. Then \eqref{R2} gives
\begin{align}\label{zetac3}
\mathcal{R}&\approx-\frac{1}{3}\ln\left(1+\frac{\delta\pi_*}{\pi_*}\right)
+\frac1{\tilde\eta}\ln\left[1+\frac{\delta\pi_*}{\pi_*+(3-\tilde\eta)\tilde\vf_*}\right].
\end{align}
The general case when these two terms are comparable is complicated.
But it can be simplified in the limiting cases when one of the terms dominates.

When $V'(\vf_*)$ is continuous, which we call a smooth transition \cite{Cai:2018dkf}, we have $\tilde\vf_*=\vf_*-\vf_m=0$ from \eqref{def:vfm}, which gives
\begin{align}
\mathcal{R}\approx\left(-\frac{1}{3}+\frac1{\tilde\eta}\right)\ln\left(1+\frac{\delta\pi_*}{\pi_*}\right)
\approx\frac1{\tilde\eta}\ln\left(1+\frac{3\delta\vf}{\pi_*}\right).
\end{align}
We see that it is similar to the slow-roll result \eqref{Rslowroll}. The only difference is the coefficient 
in front of $\delta\vf$ inside the logarithm. 
This means that in a smooth transition $\calR$ is dominated by the contribution from the
second slow-roll stage, which generates the same perturbation series as \eqref{Rseries}
with $\calR_g=(3/\tilde\eta)(\delta\vf/\pi_*)$ and $f_\text{NL}=-5\tilde\eta/6$. 

The opposite limit is when the discontinuity in $V'(\vf_*)$ is large,  
i.e., $(3-\tilde\eta)|\vf_*-\vf_m|\gg\pi_*$, which we call a sharp transition \cite{Cai:2018dkf,Passaglia:2018ixg}. 
Now the $\delta\pi_*$-term in the second logarithm of \eqref{zetac3} is much suppressed and always negligible compared to the first term, yielding
\begin{align}\label{Rusr}
\mathcal{R}&\approx-\frac{1}{3}\ln\left(1+\frac{\delta\pi_*}{\pi_*}\right).
\end{align}
Thus the USR result \eqref{USR} is recovered in this limit.

Recently several papers on inflation with a bumpy potential have appeared \cite{Atal:2019cdz,Atal:2019erb}.
To realize such a case, we assume that inflation is already in the attractor regime at $\vf_*$ with $m_1^2<0$. 
As we commented, the total $\delta N$ is approximated by \eqref{R4}.
A non-vanishing positive field velocity $\pi$ in the denominator is necessary 
if the inflaton comes from the other side ($\vf<0$) of the bump. 
This means $\pi$ must deviate from the attractor solution, $\pi=-\lambda_-\vf$, in the vicinity of the top of the bump, 
as is shown clearly in the phase portrait in Fig.\ref{f:phase}. 
Taking into account the conservation of $(\delta\pi+\lambda_+\delta\vf)/(\pi+\lambda_+\lambda\vf)$ on superhorizon scales, our result is in agreement with Refs. \cite{Atal:2019cdz,Atal:2019erb}.

Besides the above examples, we find some interesting new cases in which the $e$-folding number are analytically solvable. This becomes possible if the conservation law \eqref{law} can be algebraically 
solved for $\pi_*$. 
As we discussed, it can be easily solved in the USR case ($\lambda_+=3$, $\lambda_-=0$).
The other algebraically solvable cases require that $\lambda_+/\lambda_-=m/n$, where $m,~n\in\mathbb{Z}$, $m>|n|$, and $\mathrm{max}(m,m-n)\leq4$.
For instance, we have $\lambda_-=-3/2$ and $\lambda_+/\lambda_-=-3$ for $\eta=-9/4$, so 
\eqref{com5} and \eqref{com6} gives a fourth order algebraic equation for $\pi_*$,
\be\label{4th}
\left(\pi_*+\frac92\vf_*\right)^3\left(\pi_*-\frac32\vf_*\right)=\left(\pi+\frac92\vf\right)^3\left(\pi-\frac32\vf\right).
\ee
This is algebraically solvable, so $\calR(\delta\vf, \delta\pi)$ given by \eqref{main} has an analytical (though complicated) expression even if the inflaton is not in the attractor regime on the boundaries. Interestingly, \eqref{4th} has the similar form as the algebraic equation derived in the curvaton scenario \cite{Pi:2021dft}. All similar analytically solvable cases are listed in Table \ref{tab:algebra}. However, we did not consider the case of the degenerate characteristic roots $\lambda_-=\lambda_+=3/2$ nor the complex characteristic roots $\lambda_\pm=(3/2)\pm i\sqrt{3\eta-9/4}$ with $\eta>3/4$. We will leave studies of these situations for future work.

\begin{table}
 \begin{tabular}{c c c c c} 
  \hline
Order & $\eta$ & $\lambda_-$ & $\lambda_+$ & PDF~$\propto$ \\
 \hline
slow-roll & $\eta$ & $\eta$ & $3-\eta$ & Gaussian  \\ 
USR & $0$ & $0$ & 3 & $\exp(-3\calR)$   \\
3 & $-6$ & $-3$ & $6$ & $\exp(-3\calR)$   \\
4 & $-9/4$ & $-3/2$ & $9/2$ & $\exp(-3\calR/2)$   \\
4  & $12/25$ & $3/5$ & $12/5$ & $\exp(-c^2e^{6\calR/5})$   \\
3  & $9/16$ & $3/4$ & $9/4$ & $\exp(-c^2e^{3\calR/2})$   \\
2  & $2/3$ & $1$ & $2$ & $\exp(-c^2e^{2\calR})$   \\
3 & $18/25$ & $6/5$ & $9/5$ & $\exp(-c^2e^{12\calR/5})$   \\
4 & $36/49$ & $9/7$ & $12/7$ & $\exp(-c^2e^{18\calR/7})$   \\
\hline
\end{tabular}
\caption{Special cases when $\calR$ can be analytically derived. The number in the left-most column indicates the order of the algebraic equation to be solved. The right-most column shows the tail behavior of the PDF for $\calR>0$ when the inflaton is already in the attractor regime at $\vf_*$, where $c^2$'s are model-dependent coefficients.}\label{tab:algebra}
\end{table}



\textit{Discussion.}---
In this paper we studied single-field inflation with a piecewise quadratic potential, and calculated the curvature perturbation $\calR$ by using the $\delta N$ formalism. We found logarithms universally appear in the expression for $\calR$, 
and two seemingly different expressions involving logarithms from each segment of the quadratic potential are equivalent to each other, as given by \eqref{deltaN1} and \eqref{deltaN2}. We call this equivalence the \textit{logarithmic duality}.
Although we focused on the two-stage case, it is straightforward to generalize
our result to potentials with more quadratic pieces.


The total curvature perturbation $\calR$ is the sum of such logarithms from all stages. However, in the case when the inflaton is already in the attractor regime at the first boundary, the contributions to $\delta N$ from the later stages are negligible because the trajectory is unique afterwards, leaving a single logarithm of the local field perturbation, \eqref{R4}. Otherwise, if the non-attractor solution is still important on the boundary, like in the USR case, the boundary term can contribute or even dominate the curvature perturbation. 



When one of the logarithms dominates, the PDF of $\calR$ can be calculated easily 
from the Gaussian PDF of $\delta\vf$. Taking \eqref{R4} as an example, we obtain
\begin{align}\label{PDF}
P(\mathcal{R})
&=\frac{\left|\lambda_-\varphi\right| e^{\lambda_-\mathcal{R}}}{\sqrt{2\pi}\sigma_{\delta\varphi}}\exp\left[-\frac{\varphi^2\left(e^{\lambda_-\mathcal{R}}-1\right)^2}{2\sigma_{\delta\varphi}^2}\right],
\end{align}
where $\sigma_{\delta\vf}$ is the root-mean-square of $\delta\vf$. 
If $\lambda_-<0$, the PDF of $\calR$ 
has an exponential tail $\sim e^{\lambda_-\calR}$ for $\calR>0$. 
In the sharp-ended USR case, $\lambda_-=0$, we should use the dual expression \eqref{Rusr}, which gives $P(\calR)\sim e^{-3\calR}$ \cite{Biagetti:2021eep}. For positive $\lambda_-$, the suppression by the second exponent in \eqref{PDF} becomes important, which displays a Gumbel-distribution-like tail $P(\calR)\sim\exp(-c^2 e^{2\lambda_-\calR})$. 


The PBH formation is very sensitive to the tail of $P(\calR)$. 
Recently, various groups have considered the PBH formation for exponential-tail 
PDFs \cite{Biagetti:2021eep,Kitajima:2021fpq,Ferrante:2022mui,Gow:2022jfb}. 
It was found that the amplitude, central mass, as well as the shape of the PBH mass function changes significantly even in the simple single-logarithm case. On the other hand, the induced GWs are believed to be only mildly dependent on
non-Gaussianities, though only perturbative calculations have been done so far \cite{Garcia-Bellido:2016dkw,Nakama:2016gzw,Cai:2018dig,Unal:2018yaa,Adshead:2021hnm,Garcia-Saenz:2022tzu,Abe:2022xur}.
We may find a profound effect in the PBH formation and induced GWs when plural logarithms equally contribute to the curvature perturbation.

The exponential tail we found here is analogous to the tail found in the stochastic $\delta N$ formalism based on stochastic inflation~\cite{Starobinsky:1986fx,Starobinsky:1982ee,Starobinsky:1994bd}.
For instance, in Ref. \cite{Pattison:2021oen}, the effect of quantum diffusion was
studied in detail when there is an intermediate USR stage, which should coincide with our result in the drift-dominated limit. 
Unfortunately, two exponents seem to differ from each other. We suspect that the ``absorbing boundary condition'' adopted in the stochastic $\delta N$ formalism in \cite{Pattison:2021oen} cannot reflect how the USR stage ends, which is crucial in determining the final $\delta N$.
This is an interesting issue to be resolved in the future. 


\textit{Acknowledgement.}---
We would like to thank Albert Escriv\`a, Jaume Garriga, Vincent Vennin, and David Wands for useful discussions. SP thanks the hospitality of Yukawa Institute for Theoretical Physics, Kyoto University during his visit when this paper is finalized. SP is supported by the National Key Research and Development Program of China Grant No. 2021YFC2203004, by the CAS Project for Young Scientists in Basic Research YSBR-006, and by Project 12047503 of the National Natural Science Foundation of China. This work is also supported by JSPS Grant-in-Aid for Early-Career Scientists No. JP20K14461 (SP), by JSPS KAKENHI grants 19H01895, 20H04727, 20H05853 (MS), and by the World Premier International Research Center Initiative (WPI Initiative), MEXT, Japan. 

\bibliography{ref}

\begin{thebibliography}{74}%
\makeatletter
\providecommand \@ifxundefined [1]{%
 \@ifx{#1\undefined}
}%
\providecommand \@ifnum [1]{%
 \ifnum #1\expandafter \@firstoftwo
 \else \expandafter \@secondoftwo
 \fi
}%
\providecommand \@ifx [1]{%
 \ifx #1\expandafter \@firstoftwo
 \else \expandafter \@secondoftwo
 \fi
}%
\providecommand \natexlab [1]{#1}%
\providecommand \enquote  [1]{``#1''}%
\providecommand \bibnamefont  [1]{#1}%
\providecommand \bibfnamefont [1]{#1}%
\providecommand \citenamefont [1]{#1}%
\providecommand \href@noop [0]{\@secondoftwo}%
\providecommand \href [0]{\begingroup \@sanitize@url \@href}%
\providecommand \@href[1]{\@@startlink{#1}\@@href}%
\providecommand \@@href[1]{\endgroup#1\@@endlink}%
\providecommand \@sanitize@url [0]{\catcode `\\12\catcode `\$12\catcode
  `\&12\catcode `\#12\catcode `\^12\catcode `\_12\catcode `\%12\relax}%
\providecommand \@@startlink[1]{}%
\providecommand \@@endlink[0]{}%
\providecommand \url  [0]{\begingroup\@sanitize@url \@url }%
\providecommand \@url [1]{\endgroup\@href {#1}{\urlprefix }}%
\providecommand \urlprefix  [0]{URL }%
\providecommand \Eprint [0]{\href }%
\providecommand \doibase [0]{http://dx.doi.org/}%
\providecommand \selectlanguage [0]{\@gobble}%
\providecommand \bibinfo  [0]{\@secondoftwo}%
\providecommand \bibfield  [0]{\@secondoftwo}%
\providecommand \translation [1]{[#1]}%
\providecommand \BibitemOpen [0]{}%
\providecommand \bibitemStop [0]{}%
\providecommand \bibitemNoStop [0]{.\EOS\space}%
\providecommand \EOS [0]{\spacefactor3000\relax}%
\providecommand \BibitemShut  [1]{\csname bibitem#1\endcsname}%
\let\auto@bib@innerbib\@empty
\bibitem [{\citenamefont {Brout}\ \emph {et~al.}(1978)\citenamefont {Brout},
  \citenamefont {Englert},\ and\ \citenamefont {Gunzig}}]{Brout:1977ix}%
  \BibitemOpen
  \bibfield  {author} {\bibinfo {author} {\bibfnamefont {R.}~\bibnamefont
  {Brout}}, \bibinfo {author} {\bibfnamefont {F.}~\bibnamefont {Englert}}, \
  and\ \bibinfo {author} {\bibfnamefont {E.}~\bibnamefont {Gunzig}},\ }\href
  {\doibase 10.1016/0003-4916(78)90176-8} {\bibfield  {journal} {\bibinfo
  {journal} {Annals Phys.}\ }\textbf {\bibinfo {volume} {115}},\ \bibinfo
  {pages} {78} (\bibinfo {year} {1978})}\BibitemShut {NoStop}%
\bibitem [{\citenamefont {Guth}(1981)}]{Guth:1980zm}%
  \BibitemOpen
  \bibfield  {author} {\bibinfo {author} {\bibfnamefont {A.~H.}\ \bibnamefont
  {Guth}},\ }\href {\doibase 10.1103/PhysRevD.23.347} {\bibfield  {journal}
  {\bibinfo  {journal} {Phys. Rev.}\ }\textbf {\bibinfo {volume} {D23}},\
  \bibinfo {pages} {347} (\bibinfo {year} {1981})}\BibitemShut {NoStop}%
\bibitem [{\citenamefont {Starobinsky}(1980)}]{Starobinsky:1980te}%
  \BibitemOpen
  \bibfield  {author} {\bibinfo {author} {\bibfnamefont {A.~A.}\ \bibnamefont
  {Starobinsky}},\ }\href {\doibase 10.1016/0370-2693(80)90670-X} {\bibfield
  {journal} {\bibinfo  {journal} {Phys. Lett.}\ }\textbf {\bibinfo {volume}
  {91B}},\ \bibinfo {pages} {99} (\bibinfo {year} {1980})},\ \bibinfo {note}
  {[Adv. Ser. Astrophys. Cosmol.3,130(1987)]}\BibitemShut {NoStop}%
\bibitem [{\citenamefont {Mukhanov}\ and\ \citenamefont
  {Chibisov}(1981)}]{Mukhanov:1981xt}%
  \BibitemOpen
  \bibfield  {author} {\bibinfo {author} {\bibfnamefont {V.~F.}\ \bibnamefont
  {Mukhanov}}\ and\ \bibinfo {author} {\bibfnamefont {G.~V.}\ \bibnamefont
  {Chibisov}},\ }\href@noop {} {\bibfield  {journal} {\bibinfo  {journal} {JETP
  Lett.}\ }\textbf {\bibinfo {volume} {33}},\ \bibinfo {pages} {532} (\bibinfo
  {year} {1981})},\ \bibinfo {note} {[Pisma Zh. Eksp. Teor.
  Fiz.33,549(1981)]}\BibitemShut {NoStop}%
\bibitem [{\citenamefont {Linde}(1982)}]{Linde:1981mu}%
  \BibitemOpen
  \bibfield  {author} {\bibinfo {author} {\bibfnamefont {A.~D.}\ \bibnamefont
  {Linde}},\ }\href {\doibase 10.1016/0370-2693(82)91219-9} {\bibfield
  {journal} {\bibinfo  {journal} {Phys. Lett. B}\ }\textbf {\bibinfo {volume}
  {108}},\ \bibinfo {pages} {389} (\bibinfo {year} {1982})}\BibitemShut
  {NoStop}%
\bibitem [{\citenamefont {Albrecht}\ and\ \citenamefont
  {Steinhardt}(1982)}]{Albrecht:1982wi}%
  \BibitemOpen
  \bibfield  {author} {\bibinfo {author} {\bibfnamefont {A.}~\bibnamefont
  {Albrecht}}\ and\ \bibinfo {author} {\bibfnamefont {P.~J.}\ \bibnamefont
  {Steinhardt}},\ }\href {\doibase 10.1103/PhysRevLett.48.1220} {\bibfield
  {journal} {\bibinfo  {journal} {Phys. Rev. Lett.}\ }\textbf {\bibinfo
  {volume} {48}},\ \bibinfo {pages} {1220} (\bibinfo {year}
  {1982})}\BibitemShut {NoStop}%
\bibitem [{\citenamefont {Mukhanov}(1985)}]{Mukhanov:1985rz}%
  \BibitemOpen
  \bibfield  {author} {\bibinfo {author} {\bibfnamefont {V.~F.}\ \bibnamefont
  {Mukhanov}},\ }\href@noop {} {\bibfield  {journal} {\bibinfo  {journal} {JETP
  Lett.}\ }\textbf {\bibinfo {volume} {41}},\ \bibinfo {pages} {493} (\bibinfo
  {year} {1985})}\BibitemShut {NoStop}%
\bibitem [{\citenamefont {Sasaki}(1986)}]{Sasaki:1986hm}%
  \BibitemOpen
  \bibfield  {author} {\bibinfo {author} {\bibfnamefont {M.}~\bibnamefont
  {Sasaki}},\ }\href {\doibase 10.1143/PTP.76.1036} {\bibfield  {journal}
  {\bibinfo  {journal} {Prog. Theor. Phys.}\ }\textbf {\bibinfo {volume}
  {76}},\ \bibinfo {pages} {1036} (\bibinfo {year} {1986})}\BibitemShut
  {NoStop}%
\bibitem [{\citenamefont {Akrami}\ \emph {et~al.}(2020)\citenamefont {Akrami}
  \emph {et~al.}}]{Planck:2018jri}%
  \BibitemOpen
  \bibfield  {author} {\bibinfo {author} {\bibfnamefont {Y.}~\bibnamefont
  {Akrami}} \emph {et~al.} (\bibinfo {collaboration} {Planck}),\ }\href
  {\doibase 10.1051/0004-6361/201833887} {\bibfield  {journal} {\bibinfo
  {journal} {Astron. Astrophys.}\ }\textbf {\bibinfo {volume} {641}},\ \bibinfo
  {pages} {A10} (\bibinfo {year} {2020})},\ \Eprint
  {http://arxiv.org/abs/1807.06211} {arXiv:1807.06211 [astro-ph.CO]}
  \BibitemShut {NoStop}%
\bibitem [{\citenamefont {Abbott}\ \emph {et~al.}(2022)\citenamefont {Abbott}
  \emph {et~al.}}]{DES:2021wwk}%
  \BibitemOpen
  \bibfield  {author} {\bibinfo {author} {\bibfnamefont {T.~M.~C.}\
  \bibnamefont {Abbott}} \emph {et~al.} (\bibinfo {collaboration} {DES}),\
  }\href {\doibase 10.1103/PhysRevD.105.023520} {\bibfield  {journal} {\bibinfo
   {journal} {Phys. Rev. D}\ }\textbf {\bibinfo {volume} {105}},\ \bibinfo
  {pages} {023520} (\bibinfo {year} {2022})},\ \Eprint
  {http://arxiv.org/abs/2105.13549} {arXiv:2105.13549 [astro-ph.CO]}
  \BibitemShut {NoStop}%
\bibitem [{\citenamefont {Zel'dovich}(1967)}]{Zeldovich:1967lct}%
  \BibitemOpen
  \bibfield  {author} {\bibinfo {author} {\bibfnamefont {I.~D.}\ \bibnamefont
  {Zel'dovich}, \bibfnamefont {Ya.B.;~Novikov}},\ }\href@noop {} {\bibfield
  {journal} {\bibinfo  {journal} {Soviet Astron. AJ (Engl. Transl. ),}\
  }\textbf {\bibinfo {volume} {10}},\ \bibinfo {pages} {602} (\bibinfo {year}
  {1967})}\BibitemShut {NoStop}%
\bibitem [{\citenamefont {Hawking}(1971)}]{Hawking:1971ei}%
  \BibitemOpen
  \bibfield  {author} {\bibinfo {author} {\bibfnamefont {S.}~\bibnamefont
  {Hawking}},\ }\href@noop {} {\bibfield  {journal} {\bibinfo  {journal} {Mon.
  Not. Roy. Astron. Soc.}\ }\textbf {\bibinfo {volume} {152}},\ \bibinfo
  {pages} {75} (\bibinfo {year} {1971})}\BibitemShut {NoStop}%
\bibitem [{\citenamefont {Carr}\ and\ \citenamefont
  {Hawking}(1974)}]{Carr:1974nx}%
  \BibitemOpen
  \bibfield  {author} {\bibinfo {author} {\bibfnamefont {B.~J.}\ \bibnamefont
  {Carr}}\ and\ \bibinfo {author} {\bibfnamefont {S.}~\bibnamefont {Hawking}},\
  }\href@noop {} {\bibfield  {journal} {\bibinfo  {journal} {Mon. Not. Roy.
  Astron. Soc.}\ }\textbf {\bibinfo {volume} {168}},\ \bibinfo {pages} {399}
  (\bibinfo {year} {1974})}\BibitemShut {NoStop}%
\bibitem [{\citenamefont {Meszaros}(1974)}]{Meszaros:1974tb}%
  \BibitemOpen
  \bibfield  {author} {\bibinfo {author} {\bibfnamefont {P.}~\bibnamefont
  {Meszaros}},\ }\href@noop {} {\bibfield  {journal} {\bibinfo  {journal}
  {Astron. Astrophys.}\ }\textbf {\bibinfo {volume} {37}},\ \bibinfo {pages}
  {225} (\bibinfo {year} {1974})}\BibitemShut {NoStop}%
\bibitem [{\citenamefont {Carr}(1975)}]{Carr:1975qj}%
  \BibitemOpen
  \bibfield  {author} {\bibinfo {author} {\bibfnamefont {B.~J.}\ \bibnamefont
  {Carr}},\ }\href {\doibase 10.1086/153853} {\bibfield  {journal} {\bibinfo
  {journal} {Astrophys. J.}\ }\textbf {\bibinfo {volume} {201}},\ \bibinfo
  {pages} {1} (\bibinfo {year} {1975})}\BibitemShut {NoStop}%
\bibitem [{\citenamefont {Khlopov}\ \emph {et~al.}(1985)\citenamefont
  {Khlopov}, \citenamefont {Malomed},\ and\ \citenamefont
  {Zeldovich}}]{Khlopov:1985jw}%
  \BibitemOpen
  \bibfield  {author} {\bibinfo {author} {\bibfnamefont {M.}~\bibnamefont
  {Khlopov}}, \bibinfo {author} {\bibfnamefont {B.}~\bibnamefont {Malomed}}, \
  and\ \bibinfo {author} {\bibfnamefont {I.}~\bibnamefont {Zeldovich}},\
  }\href@noop {} {\bibfield  {journal} {\bibinfo  {journal} {Mon. Not. Roy.
  Astron. Soc.}\ }\textbf {\bibinfo {volume} {215}},\ \bibinfo {pages} {575}
  (\bibinfo {year} {1985})}\BibitemShut {NoStop}%
\bibitem [{\citenamefont {Matarrese}\ \emph {et~al.}(1993)\citenamefont
  {Matarrese}, \citenamefont {Pantano},\ and\ \citenamefont
  {Saez}}]{Matarrese:1992rp}%
  \BibitemOpen
  \bibfield  {author} {\bibinfo {author} {\bibfnamefont {S.}~\bibnamefont
  {Matarrese}}, \bibinfo {author} {\bibfnamefont {O.}~\bibnamefont {Pantano}},
  \ and\ \bibinfo {author} {\bibfnamefont {D.}~\bibnamefont {Saez}},\ }\href
  {\doibase 10.1103/PhysRevD.47.1311} {\bibfield  {journal} {\bibinfo
  {journal} {Phys. Rev. D}\ }\textbf {\bibinfo {volume} {47}},\ \bibinfo
  {pages} {1311} (\bibinfo {year} {1993})}\BibitemShut {NoStop}%
\bibitem [{\citenamefont {Matarrese}\ \emph {et~al.}(1994)\citenamefont
  {Matarrese}, \citenamefont {Pantano},\ and\ \citenamefont
  {Saez}}]{Matarrese:1993zf}%
  \BibitemOpen
  \bibfield  {author} {\bibinfo {author} {\bibfnamefont {S.}~\bibnamefont
  {Matarrese}}, \bibinfo {author} {\bibfnamefont {O.}~\bibnamefont {Pantano}},
  \ and\ \bibinfo {author} {\bibfnamefont {D.}~\bibnamefont {Saez}},\ }\href
  {\doibase 10.1103/PhysRevLett.72.320} {\bibfield  {journal} {\bibinfo
  {journal} {Phys. Rev. Lett.}\ }\textbf {\bibinfo {volume} {72}},\ \bibinfo
  {pages} {320} (\bibinfo {year} {1994})},\ \Eprint
  {http://arxiv.org/abs/astro-ph/9310036} {arXiv:astro-ph/9310036} \BibitemShut
  {NoStop}%
\bibitem [{\citenamefont {Matarrese}\ \emph {et~al.}(1998)\citenamefont
  {Matarrese}, \citenamefont {Mollerach},\ and\ \citenamefont
  {Bruni}}]{Matarrese:1997ay}%
  \BibitemOpen
  \bibfield  {author} {\bibinfo {author} {\bibfnamefont {S.}~\bibnamefont
  {Matarrese}}, \bibinfo {author} {\bibfnamefont {S.}~\bibnamefont
  {Mollerach}}, \ and\ \bibinfo {author} {\bibfnamefont {M.}~\bibnamefont
  {Bruni}},\ }\href {\doibase 10.1103/PhysRevD.58.043504} {\bibfield  {journal}
  {\bibinfo  {journal} {Phys. Rev. D}\ }\textbf {\bibinfo {volume} {58}},\
  \bibinfo {pages} {043504} (\bibinfo {year} {1998})},\ \Eprint
  {http://arxiv.org/abs/astro-ph/9707278} {arXiv:astro-ph/9707278} \BibitemShut
  {NoStop}%
\bibitem [{\citenamefont {Noh}\ and\ \citenamefont {Hwang}(2004)}]{Noh:2004bc}%
  \BibitemOpen
  \bibfield  {author} {\bibinfo {author} {\bibfnamefont {H.}~\bibnamefont
  {Noh}}\ and\ \bibinfo {author} {\bibfnamefont {J.-c.}\ \bibnamefont
  {Hwang}},\ }\href {\doibase 10.1103/PhysRevD.69.104011} {\bibfield  {journal}
  {\bibinfo  {journal} {Phys. Rev. D}\ }\textbf {\bibinfo {volume} {69}},\
  \bibinfo {pages} {104011} (\bibinfo {year} {2004})}\BibitemShut {NoStop}%
\bibitem [{\citenamefont {Carbone}\ and\ \citenamefont
  {Matarrese}(2005)}]{Carbone:2004iv}%
  \BibitemOpen
  \bibfield  {author} {\bibinfo {author} {\bibfnamefont {C.}~\bibnamefont
  {Carbone}}\ and\ \bibinfo {author} {\bibfnamefont {S.}~\bibnamefont
  {Matarrese}},\ }\href {\doibase 10.1103/PhysRevD.71.043508} {\bibfield
  {journal} {\bibinfo  {journal} {Phys. Rev. D}\ }\textbf {\bibinfo {volume}
  {71}},\ \bibinfo {pages} {043508} (\bibinfo {year} {2005})},\ \Eprint
  {http://arxiv.org/abs/astro-ph/0407611} {arXiv:astro-ph/0407611} \BibitemShut
  {NoStop}%
\bibitem [{\citenamefont {Nakamura}(2007)}]{Nakamura:2004rm}%
  \BibitemOpen
  \bibfield  {author} {\bibinfo {author} {\bibfnamefont {K.}~\bibnamefont
  {Nakamura}},\ }\href {\doibase 10.1143/PTP.117.17} {\bibfield  {journal}
  {\bibinfo  {journal} {Prog. Theor. Phys.}\ }\textbf {\bibinfo {volume}
  {117}},\ \bibinfo {pages} {17} (\bibinfo {year} {2007})},\ \Eprint
  {http://arxiv.org/abs/gr-qc/0605108} {arXiv:gr-qc/0605108} \BibitemShut
  {NoStop}%
\bibitem [{\citenamefont {Ananda}\ \emph {et~al.}(2007)\citenamefont {Ananda},
  \citenamefont {Clarkson},\ and\ \citenamefont {Wands}}]{Ananda:2006af}%
  \BibitemOpen
  \bibfield  {author} {\bibinfo {author} {\bibfnamefont {K.~N.}\ \bibnamefont
  {Ananda}}, \bibinfo {author} {\bibfnamefont {C.}~\bibnamefont {Clarkson}}, \
  and\ \bibinfo {author} {\bibfnamefont {D.}~\bibnamefont {Wands}},\ }\href
  {\doibase 10.1103/PhysRevD.75.123518} {\bibfield  {journal} {\bibinfo
  {journal} {Phys. Rev. D}\ }\textbf {\bibinfo {volume} {75}},\ \bibinfo
  {pages} {123518} (\bibinfo {year} {2007})},\ \Eprint
  {http://arxiv.org/abs/gr-qc/0612013} {arXiv:gr-qc/0612013} \BibitemShut
  {NoStop}%
\bibitem [{\citenamefont {Baumann}\ \emph {et~al.}(2007)\citenamefont
  {Baumann}, \citenamefont {Steinhardt}, \citenamefont {Takahashi},\ and\
  \citenamefont {Ichiki}}]{Baumann:2007zm}%
  \BibitemOpen
  \bibfield  {author} {\bibinfo {author} {\bibfnamefont {D.}~\bibnamefont
  {Baumann}}, \bibinfo {author} {\bibfnamefont {P.~J.}\ \bibnamefont
  {Steinhardt}}, \bibinfo {author} {\bibfnamefont {K.}~\bibnamefont
  {Takahashi}}, \ and\ \bibinfo {author} {\bibfnamefont {K.}~\bibnamefont
  {Ichiki}},\ }\href {\doibase 10.1103/PhysRevD.76.084019} {\bibfield
  {journal} {\bibinfo  {journal} {Phys. Rev. D}\ }\textbf {\bibinfo {volume}
  {76}},\ \bibinfo {pages} {084019} (\bibinfo {year} {2007})},\ \Eprint
  {http://arxiv.org/abs/hep-th/0703290} {arXiv:hep-th/0703290} \BibitemShut
  {NoStop}%
\bibitem [{\citenamefont {Sasaki}\ and\ \citenamefont
  {Stewart}(1996)}]{Sasaki:1995aw}%
  \BibitemOpen
  \bibfield  {author} {\bibinfo {author} {\bibfnamefont {M.}~\bibnamefont
  {Sasaki}}\ and\ \bibinfo {author} {\bibfnamefont {E.~D.}\ \bibnamefont
  {Stewart}},\ }\href {\doibase 10.1143/PTP.95.71} {\bibfield  {journal}
  {\bibinfo  {journal} {Prog. Theor. Phys.}\ }\textbf {\bibinfo {volume}
  {95}},\ \bibinfo {pages} {71} (\bibinfo {year} {1996})},\ \Eprint
  {http://arxiv.org/abs/astro-ph/9507001} {arXiv:astro-ph/9507001} \BibitemShut
  {NoStop}%
\bibitem [{\citenamefont {Wands}\ \emph {et~al.}(2000)\citenamefont {Wands},
  \citenamefont {Malik}, \citenamefont {Lyth},\ and\ \citenamefont
  {Liddle}}]{Wands:2000dp}%
  \BibitemOpen
  \bibfield  {author} {\bibinfo {author} {\bibfnamefont {D.}~\bibnamefont
  {Wands}}, \bibinfo {author} {\bibfnamefont {K.~A.}\ \bibnamefont {Malik}},
  \bibinfo {author} {\bibfnamefont {D.~H.}\ \bibnamefont {Lyth}}, \ and\
  \bibinfo {author} {\bibfnamefont {A.~R.}\ \bibnamefont {Liddle}},\ }\href
  {\doibase 10.1103/PhysRevD.62.043527} {\bibfield  {journal} {\bibinfo
  {journal} {Phys. Rev. D}\ }\textbf {\bibinfo {volume} {62}},\ \bibinfo
  {pages} {043527} (\bibinfo {year} {2000})},\ \Eprint
  {http://arxiv.org/abs/astro-ph/0003278} {arXiv:astro-ph/0003278} \BibitemShut
  {NoStop}%
\bibitem [{\citenamefont {Lyth}\ \emph {et~al.}(2005)\citenamefont {Lyth},
  \citenamefont {Malik},\ and\ \citenamefont {Sasaki}}]{Lyth:2004gb}%
  \BibitemOpen
  \bibfield  {author} {\bibinfo {author} {\bibfnamefont {D.~H.}\ \bibnamefont
  {Lyth}}, \bibinfo {author} {\bibfnamefont {K.~A.}\ \bibnamefont {Malik}}, \
  and\ \bibinfo {author} {\bibfnamefont {M.}~\bibnamefont {Sasaki}},\ }\href
  {\doibase 10.1088/1475-7516/2005/05/004} {\bibfield  {journal} {\bibinfo
  {journal} {JCAP}\ }\textbf {\bibinfo {volume} {05}},\ \bibinfo {pages} {004}
  (\bibinfo {year} {2005})},\ \Eprint {http://arxiv.org/abs/astro-ph/0411220}
  {arXiv:astro-ph/0411220} \BibitemShut {NoStop}%
\bibitem [{\citenamefont {Lyth}\ and\ \citenamefont
  {Rodriguez}(2005)}]{Lyth:2005fi}%
  \BibitemOpen
  \bibfield  {author} {\bibinfo {author} {\bibfnamefont {D.~H.}\ \bibnamefont
  {Lyth}}\ and\ \bibinfo {author} {\bibfnamefont {Y.}~\bibnamefont
  {Rodriguez}},\ }\href {\doibase 10.1103/PhysRevLett.95.121302} {\bibfield
  {journal} {\bibinfo  {journal} {Phys. Rev. Lett.}\ }\textbf {\bibinfo
  {volume} {95}},\ \bibinfo {pages} {121302} (\bibinfo {year} {2005})},\
  \Eprint {http://arxiv.org/abs/astro-ph/0504045} {arXiv:astro-ph/0504045}
  \BibitemShut {NoStop}%
\bibitem [{\citenamefont {Young}\ and\ \citenamefont
  {Byrnes}(2013)}]{Young:2013oia}%
  \BibitemOpen
  \bibfield  {author} {\bibinfo {author} {\bibfnamefont {S.}~\bibnamefont
  {Young}}\ and\ \bibinfo {author} {\bibfnamefont {C.~T.}\ \bibnamefont
  {Byrnes}},\ }\href {\doibase 10.1088/1475-7516/2013/08/052} {\bibfield
  {journal} {\bibinfo  {journal} {JCAP}\ }\textbf {\bibinfo {volume} {08}},\
  \bibinfo {pages} {052} (\bibinfo {year} {2013})},\ \Eprint
  {http://arxiv.org/abs/1307.4995} {arXiv:1307.4995 [astro-ph.CO]} \BibitemShut
  {NoStop}%
\bibitem [{\citenamefont {Young}\ \emph {et~al.}(2016)\citenamefont {Young},
  \citenamefont {Regan},\ and\ \citenamefont {Byrnes}}]{Young:2015cyn}%
  \BibitemOpen
  \bibfield  {author} {\bibinfo {author} {\bibfnamefont {S.}~\bibnamefont
  {Young}}, \bibinfo {author} {\bibfnamefont {D.}~\bibnamefont {Regan}}, \ and\
  \bibinfo {author} {\bibfnamefont {C.~T.}\ \bibnamefont {Byrnes}},\ }\href
  {\doibase 10.1088/1475-7516/2016/02/029} {\bibfield  {journal} {\bibinfo
  {journal} {JCAP}\ }\textbf {\bibinfo {volume} {02}},\ \bibinfo {pages} {029}
  (\bibinfo {year} {2016})},\ \Eprint {http://arxiv.org/abs/1512.07224}
  {arXiv:1512.07224 [astro-ph.CO]} \BibitemShut {NoStop}%
\bibitem [{\citenamefont {Atal}\ and\ \citenamefont
  {Germani}(2019)}]{Atal:2018neu}%
  \BibitemOpen
  \bibfield  {author} {\bibinfo {author} {\bibfnamefont {V.}~\bibnamefont
  {Atal}}\ and\ \bibinfo {author} {\bibfnamefont {C.}~\bibnamefont {Germani}},\
  }\href {\doibase 10.1016/j.dark.2019.100275} {\bibfield  {journal} {\bibinfo
  {journal} {Phys. Dark Univ.}\ }\textbf {\bibinfo {volume} {24}},\ \bibinfo
  {pages} {100275} (\bibinfo {year} {2019})},\ \Eprint
  {http://arxiv.org/abs/1811.07857} {arXiv:1811.07857 [astro-ph.CO]}
  \BibitemShut {NoStop}%
\bibitem [{\citenamefont {Passaglia}\ \emph {et~al.}(2019)\citenamefont
  {Passaglia}, \citenamefont {Hu},\ and\ \citenamefont
  {Motohashi}}]{Passaglia:2018ixg}%
  \BibitemOpen
  \bibfield  {author} {\bibinfo {author} {\bibfnamefont {S.}~\bibnamefont
  {Passaglia}}, \bibinfo {author} {\bibfnamefont {W.}~\bibnamefont {Hu}}, \
  and\ \bibinfo {author} {\bibfnamefont {H.}~\bibnamefont {Motohashi}},\ }\href
  {\doibase 10.1103/PhysRevD.99.043536} {\bibfield  {journal} {\bibinfo
  {journal} {Phys. Rev. D}\ }\textbf {\bibinfo {volume} {99}},\ \bibinfo
  {pages} {043536} (\bibinfo {year} {2019})},\ \Eprint
  {http://arxiv.org/abs/1812.08243} {arXiv:1812.08243 [astro-ph.CO]}
  \BibitemShut {NoStop}%
\bibitem [{\citenamefont {Yoo}\ \emph {et~al.}(2019)\citenamefont {Yoo},
  \citenamefont {Gong},\ and\ \citenamefont {Yokoyama}}]{Yoo:2019pma}%
  \BibitemOpen
  \bibfield  {author} {\bibinfo {author} {\bibfnamefont {C.-M.}\ \bibnamefont
  {Yoo}}, \bibinfo {author} {\bibfnamefont {J.-O.}\ \bibnamefont {Gong}}, \
  and\ \bibinfo {author} {\bibfnamefont {S.}~\bibnamefont {Yokoyama}},\ }\href
  {\doibase 10.1088/1475-7516/2019/09/033} {\bibfield  {journal} {\bibinfo
  {journal} {JCAP}\ }\textbf {\bibinfo {volume} {09}},\ \bibinfo {pages} {033}
  (\bibinfo {year} {2019})},\ \Eprint {http://arxiv.org/abs/1906.06790}
  {arXiv:1906.06790 [astro-ph.CO]} \BibitemShut {NoStop}%
\bibitem [{\citenamefont {Kehagias}\ \emph {et~al.}(2019)\citenamefont
  {Kehagias}, \citenamefont {Musco},\ and\ \citenamefont
  {Riotto}}]{Kehagias:2019eil}%
  \BibitemOpen
  \bibfield  {author} {\bibinfo {author} {\bibfnamefont {A.}~\bibnamefont
  {Kehagias}}, \bibinfo {author} {\bibfnamefont {I.}~\bibnamefont {Musco}}, \
  and\ \bibinfo {author} {\bibfnamefont {A.}~\bibnamefont {Riotto}},\ }\href
  {\doibase 10.1088/1475-7516/2019/12/029} {\bibfield  {journal} {\bibinfo
  {journal} {JCAP}\ }\textbf {\bibinfo {volume} {12}},\ \bibinfo {pages} {029}
  (\bibinfo {year} {2019})},\ \Eprint {http://arxiv.org/abs/1906.07135}
  {arXiv:1906.07135 [astro-ph.CO]} \BibitemShut {NoStop}%
\bibitem [{\citenamefont {Mahbub}(2020)}]{Mahbub:2020row}%
  \BibitemOpen
  \bibfield  {author} {\bibinfo {author} {\bibfnamefont {R.}~\bibnamefont
  {Mahbub}},\ }\href {\doibase 10.1103/PhysRevD.102.023538} {\bibfield
  {journal} {\bibinfo  {journal} {Phys. Rev. D}\ }\textbf {\bibinfo {volume}
  {102}},\ \bibinfo {pages} {023538} (\bibinfo {year} {2020})},\ \Eprint
  {http://arxiv.org/abs/2005.03618} {arXiv:2005.03618 [astro-ph.CO]}
  \BibitemShut {NoStop}%
\bibitem [{\citenamefont {Riccardi}\ \emph {et~al.}(2021)\citenamefont
  {Riccardi}, \citenamefont {Taoso},\ and\ \citenamefont
  {Urbano}}]{Riccardi:2021rlf}%
  \BibitemOpen
  \bibfield  {author} {\bibinfo {author} {\bibfnamefont {F.}~\bibnamefont
  {Riccardi}}, \bibinfo {author} {\bibfnamefont {M.}~\bibnamefont {Taoso}}, \
  and\ \bibinfo {author} {\bibfnamefont {A.}~\bibnamefont {Urbano}},\
  }\href@noop {} {\  (\bibinfo {year} {2021})},\ \Eprint
  {http://arxiv.org/abs/2102.04084} {arXiv:2102.04084 [astro-ph.CO]}
  \BibitemShut {NoStop}%
\bibitem [{\citenamefont {Davies}\ \emph {et~al.}(2022)\citenamefont {Davies},
  \citenamefont {Carrilho},\ and\ \citenamefont {Mulryne}}]{Davies:2021loj}%
  \BibitemOpen
  \bibfield  {author} {\bibinfo {author} {\bibfnamefont {M.~W.}\ \bibnamefont
  {Davies}}, \bibinfo {author} {\bibfnamefont {P.}~\bibnamefont {Carrilho}}, \
  and\ \bibinfo {author} {\bibfnamefont {D.~J.}\ \bibnamefont {Mulryne}},\
  }\href {\doibase 10.1088/1475-7516/2022/06/019} {\bibfield  {journal}
  {\bibinfo  {journal} {JCAP}\ }\textbf {\bibinfo {volume} {06}},\ \bibinfo
  {pages} {019} (\bibinfo {year} {2022})},\ \Eprint
  {http://arxiv.org/abs/2110.08189} {arXiv:2110.08189 [astro-ph.CO]}
  \BibitemShut {NoStop}%
\bibitem [{\citenamefont {Young}(2022)}]{Young:2022phe}%
  \BibitemOpen
  \bibfield  {author} {\bibinfo {author} {\bibfnamefont {S.}~\bibnamefont
  {Young}},\ }\href {\doibase 10.1088/1475-7516/2022/05/037} {\bibfield
  {journal} {\bibinfo  {journal} {JCAP}\ }\textbf {\bibinfo {volume} {05}},\
  \bibinfo {pages} {037} (\bibinfo {year} {2022})},\ \Eprint
  {http://arxiv.org/abs/2201.13345} {arXiv:2201.13345 [astro-ph.CO]}
  \BibitemShut {NoStop}%
\bibitem [{\citenamefont {Escriv\`a}\ \emph {et~al.}(2022)\citenamefont
  {Escriv\`a}, \citenamefont {Tada}, \citenamefont {Yokoyama},\ and\
  \citenamefont {Yoo}}]{Escriva:2022pnz}%
  \BibitemOpen
  \bibfield  {author} {\bibinfo {author} {\bibfnamefont {A.}~\bibnamefont
  {Escriv\`a}}, \bibinfo {author} {\bibfnamefont {Y.}~\bibnamefont {Tada}},
  \bibinfo {author} {\bibfnamefont {S.}~\bibnamefont {Yokoyama}}, \ and\
  \bibinfo {author} {\bibfnamefont {C.-M.}\ \bibnamefont {Yoo}},\ }\href
  {\doibase 10.1088/1475-7516/2022/05/012} {\bibfield  {journal} {\bibinfo
  {journal} {JCAP}\ }\textbf {\bibinfo {volume} {05}},\ \bibinfo {pages} {012}
  (\bibinfo {year} {2022})},\ \Eprint {http://arxiv.org/abs/2202.01028}
  {arXiv:2202.01028 [astro-ph.CO]} \BibitemShut {NoStop}%
\bibitem [{\citenamefont {Matsubara}\ and\ \citenamefont
  {Sasaki}(2022)}]{Matsubara:2022nbr}%
  \BibitemOpen
  \bibfield  {author} {\bibinfo {author} {\bibfnamefont {T.}~\bibnamefont
  {Matsubara}}\ and\ \bibinfo {author} {\bibfnamefont {M.}~\bibnamefont
  {Sasaki}},\ }\href {\doibase 10.1088/1475-7516/2022/10/094} {\bibfield
  {journal} {\bibinfo  {journal} {JCAP}\ }\textbf {\bibinfo {volume} {10}},\
  \bibinfo {pages} {094} (\bibinfo {year} {2022})},\ \Eprint
  {http://arxiv.org/abs/2208.02941} {arXiv:2208.02941 [astro-ph.CO]}
  \BibitemShut {NoStop}%
\bibitem [{\citenamefont {Cai}\ \emph {et~al.}(2018)\citenamefont {Cai},
  \citenamefont {Chen}, \citenamefont {Namjoo}, \citenamefont {Sasaki},
  \citenamefont {Wang},\ and\ \citenamefont {Wang}}]{Cai:2018dkf}%
  \BibitemOpen
  \bibfield  {author} {\bibinfo {author} {\bibfnamefont {Y.-F.}\ \bibnamefont
  {Cai}}, \bibinfo {author} {\bibfnamefont {X.}~\bibnamefont {Chen}}, \bibinfo
  {author} {\bibfnamefont {M.~H.}\ \bibnamefont {Namjoo}}, \bibinfo {author}
  {\bibfnamefont {M.}~\bibnamefont {Sasaki}}, \bibinfo {author} {\bibfnamefont
  {D.-G.}\ \bibnamefont {Wang}}, \ and\ \bibinfo {author} {\bibfnamefont
  {Z.}~\bibnamefont {Wang}},\ }\href {\doibase 10.1088/1475-7516/2018/05/012}
  {\bibfield  {journal} {\bibinfo  {journal} {JCAP}\ }\textbf {\bibinfo
  {volume} {05}},\ \bibinfo {pages} {012} (\bibinfo {year} {2018})},\ \Eprint
  {http://arxiv.org/abs/1712.09998} {arXiv:1712.09998 [astro-ph.CO]}
  \BibitemShut {NoStop}%
\bibitem [{\citenamefont {Biagetti}\ \emph {et~al.}(2018)\citenamefont
  {Biagetti}, \citenamefont {Franciolini}, \citenamefont {Kehagias},\ and\
  \citenamefont {Riotto}}]{Biagetti:2018pjj}%
  \BibitemOpen
  \bibfield  {author} {\bibinfo {author} {\bibfnamefont {M.}~\bibnamefont
  {Biagetti}}, \bibinfo {author} {\bibfnamefont {G.}~\bibnamefont
  {Franciolini}}, \bibinfo {author} {\bibfnamefont {A.}~\bibnamefont
  {Kehagias}}, \ and\ \bibinfo {author} {\bibfnamefont {A.}~\bibnamefont
  {Riotto}},\ }\href {\doibase 10.1088/1475-7516/2018/07/032} {\bibfield
  {journal} {\bibinfo  {journal} {JCAP}\ }\textbf {\bibinfo {volume} {07}},\
  \bibinfo {pages} {032} (\bibinfo {year} {2018})},\ \Eprint
  {http://arxiv.org/abs/1804.07124} {arXiv:1804.07124 [astro-ph.CO]}
  \BibitemShut {NoStop}%
\bibitem [{\citenamefont {Atal}\ \emph {et~al.}(2020)\citenamefont {Atal},
  \citenamefont {Cid}, \citenamefont {Escriv\`a},\ and\ \citenamefont
  {Garriga}}]{Atal:2019erb}%
  \BibitemOpen
  \bibfield  {author} {\bibinfo {author} {\bibfnamefont {V.}~\bibnamefont
  {Atal}}, \bibinfo {author} {\bibfnamefont {J.}~\bibnamefont {Cid}}, \bibinfo
  {author} {\bibfnamefont {A.}~\bibnamefont {Escriv\`a}}, \ and\ \bibinfo
  {author} {\bibfnamefont {J.}~\bibnamefont {Garriga}},\ }\href {\doibase
  10.1088/1475-7516/2020/05/022} {\bibfield  {journal} {\bibinfo  {journal}
  {JCAP}\ }\textbf {\bibinfo {volume} {05}},\ \bibinfo {pages} {022} (\bibinfo
  {year} {2020})},\ \Eprint {http://arxiv.org/abs/1908.11357} {arXiv:1908.11357
  [astro-ph.CO]} \BibitemShut {NoStop}%
\bibitem [{\citenamefont {Atal}\ \emph {et~al.}(2019)\citenamefont {Atal},
  \citenamefont {Garriga},\ and\ \citenamefont
  {Marcos-Caballero}}]{Atal:2019cdz}%
  \BibitemOpen
  \bibfield  {author} {\bibinfo {author} {\bibfnamefont {V.}~\bibnamefont
  {Atal}}, \bibinfo {author} {\bibfnamefont {J.}~\bibnamefont {Garriga}}, \
  and\ \bibinfo {author} {\bibfnamefont {A.}~\bibnamefont {Marcos-Caballero}},\
  }\href {\doibase 10.1088/1475-7516/2019/09/073} {\bibfield  {journal}
  {\bibinfo  {journal} {JCAP}\ }\textbf {\bibinfo {volume} {09}},\ \bibinfo
  {pages} {073} (\bibinfo {year} {2019})},\ \Eprint
  {http://arxiv.org/abs/1905.13202} {arXiv:1905.13202 [astro-ph.CO]}
  \BibitemShut {NoStop}%
\bibitem [{\citenamefont {Pi}\ and\ \citenamefont {Sasaki}(2021)}]{Pi:2021dft}%
  \BibitemOpen
  \bibfield  {author} {\bibinfo {author} {\bibfnamefont {S.}~\bibnamefont
  {Pi}}\ and\ \bibinfo {author} {\bibfnamefont {M.}~\bibnamefont {Sasaki}},\
  }\href@noop {} {\  (\bibinfo {year} {2021})},\ \Eprint
  {http://arxiv.org/abs/2112.12680} {arXiv:2112.12680 [astro-ph.CO]}
  \BibitemShut {NoStop}%
\bibitem [{\citenamefont {Cai}\ \emph {et~al.}(2022{\natexlab{a}})\citenamefont
  {Cai}, \citenamefont {Ma}, \citenamefont {Sasaki}, \citenamefont {Wang},\
  and\ \citenamefont {Zhou}}]{Cai:2021zsp}%
  \BibitemOpen
  \bibfield  {author} {\bibinfo {author} {\bibfnamefont {Y.-F.}\ \bibnamefont
  {Cai}}, \bibinfo {author} {\bibfnamefont {X.-H.}\ \bibnamefont {Ma}},
  \bibinfo {author} {\bibfnamefont {M.}~\bibnamefont {Sasaki}}, \bibinfo
  {author} {\bibfnamefont {D.-G.}\ \bibnamefont {Wang}}, \ and\ \bibinfo
  {author} {\bibfnamefont {Z.}~\bibnamefont {Zhou}},\ }\href {\doibase
  10.1016/j.physletb.2022.137461} {\bibfield  {journal} {\bibinfo  {journal}
  {Phys. Lett. B}\ }\textbf {\bibinfo {volume} {834}},\ \bibinfo {pages}
  {137461} (\bibinfo {year} {2022}{\natexlab{a}})},\ \Eprint
  {http://arxiv.org/abs/2112.13836} {arXiv:2112.13836 [astro-ph.CO]}
  \BibitemShut {NoStop}%
\bibitem [{\citenamefont {Cai}\ \emph {et~al.}(2022{\natexlab{b}})\citenamefont
  {Cai}, \citenamefont {Ma}, \citenamefont {Sasaki}, \citenamefont {Wang},\
  and\ \citenamefont {Zhou}}]{Cai:2022erk}%
  \BibitemOpen
  \bibfield  {author} {\bibinfo {author} {\bibfnamefont {Y.-F.}\ \bibnamefont
  {Cai}}, \bibinfo {author} {\bibfnamefont {X.-H.}\ \bibnamefont {Ma}},
  \bibinfo {author} {\bibfnamefont {M.}~\bibnamefont {Sasaki}}, \bibinfo
  {author} {\bibfnamefont {D.-G.}\ \bibnamefont {Wang}}, \ and\ \bibinfo
  {author} {\bibfnamefont {Z.}~\bibnamefont {Zhou}},\ }\href@noop {} {\
  (\bibinfo {year} {2022}{\natexlab{b}})},\ \Eprint
  {http://arxiv.org/abs/2207.11910} {arXiv:2207.11910 [astro-ph.CO]}
  \BibitemShut {NoStop}%
\bibitem [{\citenamefont {Vennin}\ and\ \citenamefont
  {Starobinsky}(2015)}]{Vennin:2015hra}%
  \BibitemOpen
  \bibfield  {author} {\bibinfo {author} {\bibfnamefont {V.}~\bibnamefont
  {Vennin}}\ and\ \bibinfo {author} {\bibfnamefont {A.~A.}\ \bibnamefont
  {Starobinsky}},\ }\href {\doibase 10.1140/epjc/s10052-015-3643-y} {\bibfield
  {journal} {\bibinfo  {journal} {Eur. Phys. J. C}\ }\textbf {\bibinfo {volume}
  {75}},\ \bibinfo {pages} {413} (\bibinfo {year} {2015})},\ \Eprint
  {http://arxiv.org/abs/1506.04732} {arXiv:1506.04732 [hep-th]} \BibitemShut
  {NoStop}%
\bibitem [{\citenamefont {Pattison}\ \emph {et~al.}(2017)\citenamefont
  {Pattison}, \citenamefont {Vennin}, \citenamefont {Assadullahi},\ and\
  \citenamefont {Wands}}]{Pattison:2017mbe}%
  \BibitemOpen
  \bibfield  {author} {\bibinfo {author} {\bibfnamefont {C.}~\bibnamefont
  {Pattison}}, \bibinfo {author} {\bibfnamefont {V.}~\bibnamefont {Vennin}},
  \bibinfo {author} {\bibfnamefont {H.}~\bibnamefont {Assadullahi}}, \ and\
  \bibinfo {author} {\bibfnamefont {D.}~\bibnamefont {Wands}},\ }\href
  {\doibase 10.1088/1475-7516/2017/10/046} {\bibfield  {journal} {\bibinfo
  {journal} {JCAP}\ }\textbf {\bibinfo {volume} {10}},\ \bibinfo {pages} {046}
  (\bibinfo {year} {2017})},\ \Eprint {http://arxiv.org/abs/1707.00537}
  {arXiv:1707.00537 [hep-th]} \BibitemShut {NoStop}%
\bibitem [{\citenamefont {Ezquiaga}\ \emph {et~al.}(2020)\citenamefont
  {Ezquiaga}, \citenamefont {Garc\'\i{}a-Bellido},\ and\ \citenamefont
  {Vennin}}]{Ezquiaga:2019ftu}%
  \BibitemOpen
  \bibfield  {author} {\bibinfo {author} {\bibfnamefont {J.~M.}\ \bibnamefont
  {Ezquiaga}}, \bibinfo {author} {\bibfnamefont {J.}~\bibnamefont
  {Garc\'\i{}a-Bellido}}, \ and\ \bibinfo {author} {\bibfnamefont
  {V.}~\bibnamefont {Vennin}},\ }\href {\doibase 10.1088/1475-7516/2020/03/029}
  {\bibfield  {journal} {\bibinfo  {journal} {JCAP}\ }\textbf {\bibinfo
  {volume} {03}},\ \bibinfo {pages} {029} (\bibinfo {year} {2020})},\ \Eprint
  {http://arxiv.org/abs/1912.05399} {arXiv:1912.05399 [astro-ph.CO]}
  \BibitemShut {NoStop}%
\bibitem [{\citenamefont {Vennin}(2020)}]{Vennin:2020kng}%
  \BibitemOpen
  \bibfield  {author} {\bibinfo {author} {\bibfnamefont {V.}~\bibnamefont
  {Vennin}},\ }\emph {\bibinfo {title} {{Stochastic inflation and primordial
  black holes}}},\ \href@noop {} {Ph.D. thesis},\ \bibinfo  {school} {U.
  Paris-Saclay} (\bibinfo {year} {2020}),\ \Eprint
  {http://arxiv.org/abs/2009.08715} {arXiv:2009.08715 [astro-ph.CO]}
  \BibitemShut {NoStop}%
\bibitem [{\citenamefont {Figueroa}\ \emph {et~al.}(2021)\citenamefont
  {Figueroa}, \citenamefont {Raatikainen}, \citenamefont {Rasanen},\ and\
  \citenamefont {Tomberg}}]{Figueroa:2020jkf}%
  \BibitemOpen
  \bibfield  {author} {\bibinfo {author} {\bibfnamefont {D.~G.}\ \bibnamefont
  {Figueroa}}, \bibinfo {author} {\bibfnamefont {S.}~\bibnamefont
  {Raatikainen}}, \bibinfo {author} {\bibfnamefont {S.}~\bibnamefont
  {Rasanen}}, \ and\ \bibinfo {author} {\bibfnamefont {E.}~\bibnamefont
  {Tomberg}},\ }\href {\doibase 10.1103/PhysRevLett.127.101302} {\bibfield
  {journal} {\bibinfo  {journal} {Phys. Rev. Lett.}\ }\textbf {\bibinfo
  {volume} {127}},\ \bibinfo {pages} {101302} (\bibinfo {year} {2021})},\
  \Eprint {http://arxiv.org/abs/2012.06551} {arXiv:2012.06551 [astro-ph.CO]}
  \BibitemShut {NoStop}%
\bibitem [{\citenamefont {Pattison}\ \emph {et~al.}(2021)\citenamefont
  {Pattison}, \citenamefont {Vennin}, \citenamefont {Wands},\ and\
  \citenamefont {Assadullahi}}]{Pattison:2021oen}%
  \BibitemOpen
  \bibfield  {author} {\bibinfo {author} {\bibfnamefont {C.}~\bibnamefont
  {Pattison}}, \bibinfo {author} {\bibfnamefont {V.}~\bibnamefont {Vennin}},
  \bibinfo {author} {\bibfnamefont {D.}~\bibnamefont {Wands}}, \ and\ \bibinfo
  {author} {\bibfnamefont {H.}~\bibnamefont {Assadullahi}},\ }\href {\doibase
  10.1088/1475-7516/2021/04/080} {\bibfield  {journal} {\bibinfo  {journal}
  {JCAP}\ }\textbf {\bibinfo {volume} {04}},\ \bibinfo {pages} {080} (\bibinfo
  {year} {2021})},\ \Eprint {http://arxiv.org/abs/2101.05741} {arXiv:2101.05741
  [astro-ph.CO]} \BibitemShut {NoStop}%
\bibitem [{\citenamefont {Figueroa}\ \emph {et~al.}(2022)\citenamefont
  {Figueroa}, \citenamefont {Raatikainen}, \citenamefont {Rasanen},\ and\
  \citenamefont {Tomberg}}]{Figueroa:2021zah}%
  \BibitemOpen
  \bibfield  {author} {\bibinfo {author} {\bibfnamefont {D.~G.}\ \bibnamefont
  {Figueroa}}, \bibinfo {author} {\bibfnamefont {S.}~\bibnamefont
  {Raatikainen}}, \bibinfo {author} {\bibfnamefont {S.}~\bibnamefont
  {Rasanen}}, \ and\ \bibinfo {author} {\bibfnamefont {E.}~\bibnamefont
  {Tomberg}},\ }\href {\doibase 10.1088/1475-7516/2022/05/027} {\bibfield
  {journal} {\bibinfo  {journal} {JCAP}\ }\textbf {\bibinfo {volume} {05}},\
  \bibinfo {pages} {027} (\bibinfo {year} {2022})},\ \Eprint
  {http://arxiv.org/abs/2111.07437} {arXiv:2111.07437 [astro-ph.CO]}
  \BibitemShut {NoStop}%
\bibitem [{\citenamefont {Animali}\ and\ \citenamefont
  {Vennin}(2022)}]{Animali:2022otk}%
  \BibitemOpen
  \bibfield  {author} {\bibinfo {author} {\bibfnamefont {C.}~\bibnamefont
  {Animali}}\ and\ \bibinfo {author} {\bibfnamefont {V.}~\bibnamefont
  {Vennin}},\ }\href@noop {} {\  (\bibinfo {year} {2022})},\ \Eprint
  {http://arxiv.org/abs/2210.03812} {arXiv:2210.03812 [astro-ph.CO]}
  \BibitemShut {NoStop}%
\bibitem [{\citenamefont {Karam}\ \emph {et~al.}(2022)\citenamefont {Karam},
  \citenamefont {Koivunen}, \citenamefont {Tomberg}, \citenamefont {Vaskonen},\
  and\ \citenamefont {Veerm\"ae}}]{Karam:2022nym}%
  \BibitemOpen
  \bibfield  {author} {\bibinfo {author} {\bibfnamefont {A.}~\bibnamefont
  {Karam}}, \bibinfo {author} {\bibfnamefont {N.}~\bibnamefont {Koivunen}},
  \bibinfo {author} {\bibfnamefont {E.}~\bibnamefont {Tomberg}}, \bibinfo
  {author} {\bibfnamefont {V.}~\bibnamefont {Vaskonen}}, \ and\ \bibinfo
  {author} {\bibfnamefont {H.}~\bibnamefont {Veerm\"ae}},\ }\href@noop {} {\
  (\bibinfo {year} {2022})},\ \Eprint {http://arxiv.org/abs/2205.13540}
  {arXiv:2205.13540 [astro-ph.CO]} \BibitemShut {NoStop}%
\bibitem [{\citenamefont {Namjoo}\ \emph {et~al.}(2013)\citenamefont {Namjoo},
  \citenamefont {Firouzjahi},\ and\ \citenamefont {Sasaki}}]{Namjoo:2012aa}%
  \BibitemOpen
  \bibfield  {author} {\bibinfo {author} {\bibfnamefont {M.~H.}\ \bibnamefont
  {Namjoo}}, \bibinfo {author} {\bibfnamefont {H.}~\bibnamefont {Firouzjahi}},
  \ and\ \bibinfo {author} {\bibfnamefont {M.}~\bibnamefont {Sasaki}},\ }\href
  {\doibase 10.1209/0295-5075/101/39001} {\bibfield  {journal} {\bibinfo
  {journal} {EPL}\ }\textbf {\bibinfo {volume} {101}},\ \bibinfo {pages}
  {39001} (\bibinfo {year} {2013})},\ \Eprint {http://arxiv.org/abs/1210.3692}
  {arXiv:1210.3692 [astro-ph.CO]} \BibitemShut {NoStop}%
\bibitem [{Note1()}]{Note1}%
  \BibitemOpen
  \bibinfo {note} {We thank Jaume Garriga for pointing this out.}\BibitemShut
  {Stop}%
\bibitem [{\citenamefont {Leach}\ \emph {et~al.}(2001)\citenamefont {Leach},
  \citenamefont {Sasaki}, \citenamefont {Wands},\ and\ \citenamefont
  {Liddle}}]{Leach:2001zf}%
  \BibitemOpen
  \bibfield  {author} {\bibinfo {author} {\bibfnamefont {S.~M.}\ \bibnamefont
  {Leach}}, \bibinfo {author} {\bibfnamefont {M.}~\bibnamefont {Sasaki}},
  \bibinfo {author} {\bibfnamefont {D.}~\bibnamefont {Wands}}, \ and\ \bibinfo
  {author} {\bibfnamefont {A.~R.}\ \bibnamefont {Liddle}},\ }\href {\doibase
  10.1103/PhysRevD.64.023512} {\bibfield  {journal} {\bibinfo  {journal} {Phys.
  Rev. D}\ }\textbf {\bibinfo {volume} {64}},\ \bibinfo {pages} {023512}
  (\bibinfo {year} {2001})},\ \Eprint {http://arxiv.org/abs/astro-ph/0101406}
  {arXiv:astro-ph/0101406} \BibitemShut {NoStop}%
\bibitem [{Note2()}]{Note2}%
  \BibitemOpen
  \bibinfo {note} {In this figure, we only consider $\varphi >0$. We will leave
  $\varphi \leq 0$ case for future work.}\BibitemShut {Stop}%
\bibitem [{\citenamefont {Biagetti}\ \emph {et~al.}(2021)\citenamefont
  {Biagetti}, \citenamefont {De~Luca}, \citenamefont {Franciolini},
  \citenamefont {Kehagias},\ and\ \citenamefont {Riotto}}]{Biagetti:2021eep}%
  \BibitemOpen
  \bibfield  {author} {\bibinfo {author} {\bibfnamefont {M.}~\bibnamefont
  {Biagetti}}, \bibinfo {author} {\bibfnamefont {V.}~\bibnamefont {De~Luca}},
  \bibinfo {author} {\bibfnamefont {G.}~\bibnamefont {Franciolini}}, \bibinfo
  {author} {\bibfnamefont {A.}~\bibnamefont {Kehagias}}, \ and\ \bibinfo
  {author} {\bibfnamefont {A.}~\bibnamefont {Riotto}},\ }\href {\doibase
  10.1016/j.physletb.2021.136602} {\bibfield  {journal} {\bibinfo  {journal}
  {Phys. Lett. B}\ }\textbf {\bibinfo {volume} {820}},\ \bibinfo {pages}
  {136602} (\bibinfo {year} {2021})},\ \Eprint
  {http://arxiv.org/abs/2105.07810} {arXiv:2105.07810 [astro-ph.CO]}
  \BibitemShut {NoStop}%
\bibitem [{\citenamefont {Kitajima}\ \emph {et~al.}(2021)\citenamefont
  {Kitajima}, \citenamefont {Tada}, \citenamefont {Yokoyama},\ and\
  \citenamefont {Yoo}}]{Kitajima:2021fpq}%
  \BibitemOpen
  \bibfield  {author} {\bibinfo {author} {\bibfnamefont {N.}~\bibnamefont
  {Kitajima}}, \bibinfo {author} {\bibfnamefont {Y.}~\bibnamefont {Tada}},
  \bibinfo {author} {\bibfnamefont {S.}~\bibnamefont {Yokoyama}}, \ and\
  \bibinfo {author} {\bibfnamefont {C.-M.}\ \bibnamefont {Yoo}},\ }\href
  {\doibase 10.1088/1475-7516/2021/10/053} {\bibfield  {journal} {\bibinfo
  {journal} {JCAP}\ }\textbf {\bibinfo {volume} {10}},\ \bibinfo {pages} {053}
  (\bibinfo {year} {2021})},\ \Eprint {http://arxiv.org/abs/2109.00791}
  {arXiv:2109.00791 [astro-ph.CO]} \BibitemShut {NoStop}%
\bibitem [{\citenamefont {Ferrante}\ \emph {et~al.}(2022)\citenamefont
  {Ferrante}, \citenamefont {Franciolini}, \citenamefont {Iovino},\ and\
  \citenamefont {Urbano}}]{Ferrante:2022mui}%
  \BibitemOpen
  \bibfield  {author} {\bibinfo {author} {\bibfnamefont {G.}~\bibnamefont
  {Ferrante}}, \bibinfo {author} {\bibfnamefont {G.}~\bibnamefont
  {Franciolini}}, \bibinfo {author} {\bibfnamefont {A.}~\bibnamefont {Iovino},
  \bibfnamefont {Junior.}}, \ and\ \bibinfo {author} {\bibfnamefont
  {A.}~\bibnamefont {Urbano}},\ }\href@noop {} {\  (\bibinfo {year} {2022})},\
  \Eprint {http://arxiv.org/abs/2211.01728} {arXiv:2211.01728 [astro-ph.CO]}
  \BibitemShut {NoStop}%
\bibitem [{\citenamefont {Gow}\ \emph {et~al.}(2022)\citenamefont {Gow},
  \citenamefont {Assadullahi}, \citenamefont {Jackson}, \citenamefont {Koyama},
  \citenamefont {Vennin},\ and\ \citenamefont {Wands}}]{Gow:2022jfb}%
  \BibitemOpen
  \bibfield  {author} {\bibinfo {author} {\bibfnamefont {A.~D.}\ \bibnamefont
  {Gow}}, \bibinfo {author} {\bibfnamefont {H.}~\bibnamefont {Assadullahi}},
  \bibinfo {author} {\bibfnamefont {J.~H.~P.}\ \bibnamefont {Jackson}},
  \bibinfo {author} {\bibfnamefont {K.}~\bibnamefont {Koyama}}, \bibinfo
  {author} {\bibfnamefont {V.}~\bibnamefont {Vennin}}, \ and\ \bibinfo {author}
  {\bibfnamefont {D.}~\bibnamefont {Wands}},\ }\href@noop {} {\  (\bibinfo
  {year} {2022})},\ \Eprint {http://arxiv.org/abs/2211.08348} {arXiv:2211.08348
  [astro-ph.CO]} \BibitemShut {NoStop}%
\bibitem [{\citenamefont {Garcia-Bellido}\ \emph {et~al.}(2016)\citenamefont
  {Garcia-Bellido}, \citenamefont {Peloso},\ and\ \citenamefont
  {Unal}}]{Garcia-Bellido:2016dkw}%
  \BibitemOpen
  \bibfield  {author} {\bibinfo {author} {\bibfnamefont {J.}~\bibnamefont
  {Garcia-Bellido}}, \bibinfo {author} {\bibfnamefont {M.}~\bibnamefont
  {Peloso}}, \ and\ \bibinfo {author} {\bibfnamefont {C.}~\bibnamefont
  {Unal}},\ }\href {\doibase 10.1088/1475-7516/2016/12/031} {\bibfield
  {journal} {\bibinfo  {journal} {JCAP}\ }\textbf {\bibinfo {volume} {12}},\
  \bibinfo {pages} {031} (\bibinfo {year} {2016})},\ \Eprint
  {http://arxiv.org/abs/1610.03763} {arXiv:1610.03763 [astro-ph.CO]}
  \BibitemShut {NoStop}%
\bibitem [{\citenamefont {Nakama}\ \emph {et~al.}(2017)\citenamefont {Nakama},
  \citenamefont {Silk},\ and\ \citenamefont {Kamionkowski}}]{Nakama:2016gzw}%
  \BibitemOpen
  \bibfield  {author} {\bibinfo {author} {\bibfnamefont {T.}~\bibnamefont
  {Nakama}}, \bibinfo {author} {\bibfnamefont {J.}~\bibnamefont {Silk}}, \ and\
  \bibinfo {author} {\bibfnamefont {M.}~\bibnamefont {Kamionkowski}},\ }\href
  {\doibase 10.1103/PhysRevD.95.043511} {\bibfield  {journal} {\bibinfo
  {journal} {Phys. Rev. D}\ }\textbf {\bibinfo {volume} {95}},\ \bibinfo
  {pages} {043511} (\bibinfo {year} {2017})},\ \Eprint
  {http://arxiv.org/abs/1612.06264} {arXiv:1612.06264 [astro-ph.CO]}
  \BibitemShut {NoStop}%
\bibitem [{\citenamefont {Cai}\ \emph {et~al.}(2019)\citenamefont {Cai},
  \citenamefont {Pi},\ and\ \citenamefont {Sasaki}}]{Cai:2018dig}%
  \BibitemOpen
  \bibfield  {author} {\bibinfo {author} {\bibfnamefont {R.-g.}\ \bibnamefont
  {Cai}}, \bibinfo {author} {\bibfnamefont {S.}~\bibnamefont {Pi}}, \ and\
  \bibinfo {author} {\bibfnamefont {M.}~\bibnamefont {Sasaki}},\ }\href
  {\doibase 10.1103/PhysRevLett.122.201101} {\bibfield  {journal} {\bibinfo
  {journal} {Phys. Rev. Lett.}\ }\textbf {\bibinfo {volume} {122}},\ \bibinfo
  {pages} {201101} (\bibinfo {year} {2019})},\ \Eprint
  {http://arxiv.org/abs/1810.11000} {arXiv:1810.11000 [astro-ph.CO]}
  \BibitemShut {NoStop}%
\bibitem [{\citenamefont {Unal}(2019)}]{Unal:2018yaa}%
  \BibitemOpen
  \bibfield  {author} {\bibinfo {author} {\bibfnamefont {C.}~\bibnamefont
  {Unal}},\ }\href {\doibase 10.1103/PhysRevD.99.041301} {\bibfield  {journal}
  {\bibinfo  {journal} {Phys. Rev. D}\ }\textbf {\bibinfo {volume} {99}},\
  \bibinfo {pages} {041301} (\bibinfo {year} {2019})},\ \Eprint
  {http://arxiv.org/abs/1811.09151} {arXiv:1811.09151 [astro-ph.CO]}
  \BibitemShut {NoStop}%
\bibitem [{\citenamefont {Adshead}\ \emph {et~al.}(2021)\citenamefont
  {Adshead}, \citenamefont {Lozanov},\ and\ \citenamefont
  {Weiner}}]{Adshead:2021hnm}%
  \BibitemOpen
  \bibfield  {author} {\bibinfo {author} {\bibfnamefont {P.}~\bibnamefont
  {Adshead}}, \bibinfo {author} {\bibfnamefont {K.~D.}\ \bibnamefont
  {Lozanov}}, \ and\ \bibinfo {author} {\bibfnamefont {Z.~J.}\ \bibnamefont
  {Weiner}},\ }\href {\doibase 10.1088/1475-7516/2021/10/080} {\bibfield
  {journal} {\bibinfo  {journal} {JCAP}\ }\textbf {\bibinfo {volume} {10}},\
  \bibinfo {pages} {080} (\bibinfo {year} {2021})},\ \Eprint
  {http://arxiv.org/abs/2105.01659} {arXiv:2105.01659 [astro-ph.CO]}
  \BibitemShut {NoStop}%
\bibitem [{\citenamefont {Garcia-Saenz}\ \emph {et~al.}(2022)\citenamefont
  {Garcia-Saenz}, \citenamefont {Pinol}, \citenamefont {Renaux-Petel},\ and\
  \citenamefont {Werth}}]{Garcia-Saenz:2022tzu}%
  \BibitemOpen
  \bibfield  {author} {\bibinfo {author} {\bibfnamefont {S.}~\bibnamefont
  {Garcia-Saenz}}, \bibinfo {author} {\bibfnamefont {L.}~\bibnamefont {Pinol}},
  \bibinfo {author} {\bibfnamefont {S.}~\bibnamefont {Renaux-Petel}}, \ and\
  \bibinfo {author} {\bibfnamefont {D.}~\bibnamefont {Werth}},\ }\href@noop {}
  {\  (\bibinfo {year} {2022})},\ \Eprint {http://arxiv.org/abs/2207.14267}
  {arXiv:2207.14267 [astro-ph.CO]} \BibitemShut {NoStop}%
\bibitem [{\citenamefont {Abe}\ \emph {et~al.}(2022)\citenamefont {Abe},
  \citenamefont {Inui}, \citenamefont {Tada},\ and\ \citenamefont
  {Yokoyama}}]{Abe:2022xur}%
  \BibitemOpen
  \bibfield  {author} {\bibinfo {author} {\bibfnamefont {K.~T.}\ \bibnamefont
  {Abe}}, \bibinfo {author} {\bibfnamefont {R.}~\bibnamefont {Inui}}, \bibinfo
  {author} {\bibfnamefont {Y.}~\bibnamefont {Tada}}, \ and\ \bibinfo {author}
  {\bibfnamefont {S.}~\bibnamefont {Yokoyama}},\ }\href@noop {} {\  (\bibinfo
  {year} {2022})},\ \Eprint {http://arxiv.org/abs/2209.13891} {arXiv:2209.13891
  [astro-ph.CO]} \BibitemShut {NoStop}%
\bibitem [{\citenamefont {Starobinsky}(1986)}]{Starobinsky:1986fx}%
  \BibitemOpen
  \bibfield  {author} {\bibinfo {author} {\bibfnamefont {A.~A.}\ \bibnamefont
  {Starobinsky}},\ }\href {\doibase 10.1007/3-540-16452-9_6} {\bibfield
  {journal} {\bibinfo  {journal} {Lect. Notes Phys.}\ }\textbf {\bibinfo
  {volume} {246}},\ \bibinfo {pages} {107} (\bibinfo {year}
  {1986})}\BibitemShut {NoStop}%
\bibitem [{\citenamefont {Starobinsky}(1982)}]{Starobinsky:1982ee}%
  \BibitemOpen
  \bibfield  {author} {\bibinfo {author} {\bibfnamefont {A.~A.}\ \bibnamefont
  {Starobinsky}},\ }\href {\doibase 10.1016/0370-2693(82)90541-X} {\bibfield
  {journal} {\bibinfo  {journal} {Phys. Lett. B}\ }\textbf {\bibinfo {volume}
  {117}},\ \bibinfo {pages} {175} (\bibinfo {year} {1982})}\BibitemShut
  {NoStop}%
\bibitem [{\citenamefont {Starobinsky}\ and\ \citenamefont
  {Yokoyama}(1994)}]{Starobinsky:1994bd}%
  \BibitemOpen
  \bibfield  {author} {\bibinfo {author} {\bibfnamefont {A.~A.}\ \bibnamefont
  {Starobinsky}}\ and\ \bibinfo {author} {\bibfnamefont {J.}~\bibnamefont
  {Yokoyama}},\ }\href {\doibase 10.1103/PhysRevD.50.6357} {\bibfield
  {journal} {\bibinfo  {journal} {Phys. Rev. D}\ }\textbf {\bibinfo {volume}
  {50}},\ \bibinfo {pages} {6357} (\bibinfo {year} {1994})},\ \Eprint
  {http://arxiv.org/abs/astro-ph/9407016} {arXiv:astro-ph/9407016} \BibitemShut
  {NoStop}%
\end{thebibliography}%

\end{document}